%% file: main.tex
\documentclass[journal]{IEEEtran}
\usepackage{amsmath}
\usepackage{tabularx}
\usepackage{comment}
\usepackage{booktabs}
\usepackage{threeparttable}
\usepackage{multirow}
\usepackage{siunitx}
\usepackage{relsize}
\usepackage[caption=false,font=normalsize,labelfont=sf,textfont=sf]{subfig}
\usepackage{cite}
\usepackage{hyphenat}
\usepackage{textcomp}
\usepackage{xcolor}
\usepackage{amssymb}
\ifCLASSINFOpdf
   \usepackage[pdftex]{graphicx}
\else

   \usepackage[dvips]{graphicx}
\fi

\usepackage[super]{nth}

\usepackage[ruled,linesnumbered,noresetcount,vlined]{algorithm2e}
\SetKwInput{KwInput}{Input}                
\SetKwInput{KwOutput}{Output}
\renewcommand{\baselinestretch}{1}
\usepackage{float}
\floatstyle{ruled}
\newfloat{model}{thp}{lop}
\floatname{model}{Model}

\usepackage{enumitem}
\newlist{abbrv}{itemize}{1}
\setlist[abbrv,1]{label=,labelwidth=0.5in,align=parleft,itemsep=0.1\baselineskip,leftmargin=!}
\usepackage{booktabs}
\usepackage{csvsimple}

\usepackage{siunitx}
\usepackage{hyperref}
\usepackage{listings}
\usepackage{xcolor}
\usepackage{tcolorbox}
\usepackage{tikz}
\usepackage{xcolor}
\usepackage{adjustbox}
\usepackage{booktabs}
\usepackage{makecell}
\usepackage{siunitx}

\lstdefinelanguage{Julia}{
  keywords={using, const, function, return, struct, mutable, for, if, else, while, end, true, false, in, begin},
  sensitive=true,
  comment=[l]{\#},
  morecomment=[s]{\#=}{=\#},
  morestring=[b]",
  morestring=[m]',
}

\begin{document}








\title{AC-Informed DC Optimal Transmission Switching Problems via Parameter Optimization}

\author{\IEEEauthorblockN{Babak Taheri and Daniel K. Molzahn}
\thanks{\noindent B. Taheri and D.K. Molzahn are with the School of Electrical
and Computer Engineering, Georgia Institute of Technology. \{btaheri6, molzahn\}@gatech.edu. Support from NSF award \#2145564.}
}

\maketitle

\begin{abstract}
Optimal Transmission Switching (OTS) problems minimize operational costs while treating both the transmission line energization statuses and generator setpoints as decision variables. The combination of nonlinearities from an AC power flow model and discrete variables associated with line statuses makes AC-OTS a computationally challenging Mixed-Integer Nonlinear Program (MINLP). To address these challenges, the DC power flow approximation is often used to obtain a DC-OTS formulation expressed as a Mixed-Integer Linear Program (MILP). However, this approximation often leads to suboptimal or infeasible switching decisions when evaluated with an AC power flow model.
This paper proposes an enhanced DC-OTS formulation that leverages techniques for training machine learning models to optimize the DC power flow model's parameters. By optimally selecting parameter values that align flows in the DC power flow model with apparent power flows---incorporating both real and reactive components---from AC Optimal Power Flow (OPF) solutions, our method more accurately captures line congestion behavior. Integrating these optimized parameters into the DC-OTS formulation significantly improves the accuracy of switching decisions and reduces discrepancies between DC-OTS and AC-OTS solutions. We compare our optimized DC-OTS model against traditional OTS approaches, including DC-OTS, Linear Programming AC (LPAC)-OTS, and Quadratic Convex (QC)-OTS. Numerical results show that switching decisions from our model yield better performance when evaluated using an AC power flow model, with up to $44\%$ cost reductions in some cases.


\end{abstract}


\input{text/nomenclature}

\input{text/intro3}

\input{text/opf}

\input{text/proposed_algorithm}

\input{text/results}

\section{Conclusion}
\label{sec:conclusion}

In this paper, we have extended our previously developed optimized DC-OPF model to OTS applications. By refining the DC-OPF parameters, we have proposed a DC power flow model tailored to OTS settings where line congestion is crucial. The implementation of this model in OTS problems yields switching decisions that more effectively reduce operational costs when assessed with AC-OPF calculations.
This application-centric approach not only validates the utility of the optimized DC-OPF model but also demonstrates its ability to improve economic efficiency. 

\textcolor{black}{Furthermore, while our current study focuses on the base topology, our previous work in \cite{taheri2023optimizing} assessed optimized parameters tailored for each topology (i.e., under $N-1$ contingency conditions). While this yields the best performance if computational resources permit, we found that training on the base topology still leads to significant improvements over traditional approaches, suggesting strong generalization properties. In future work, we intend to explore and validate these approaches for $N-1$ security analysis, either by adapting the model to compute contingency-specific parameters or by applying the base-case optimized parameters across contingencies. Moreover, extending O‑DC‑OTS to security‑constrained and uncertainty‑aware contexts—by training parameters across contingencies and renewable scenarios or by embedding chance‑constrained/robust formulations as in~\cite{han2023optimal} and surveyed in~\cite{roald2022review}—is an important direction for future work.}

Our further work also plans to evaluate the performance of the optimized parameters in other important applications such as unit commitment and infrastructure hardening. Expanding the scope of the optimized DC-OPF model to these areas could further enhance its applicability and effectiveness in power system planning and operation.

\bibliographystyle{IEEEtran}
\IEEEtriggeratref{30} 
\bibliography{blib}

\end{document}

%% file: text/nomenclature.tex
\section*{Nomenclature}
\label{sec:nomenclature}

\begin{abbrv}
    \item[$\mathcal{N}$] Set of buses
    \item[$\mathcal{E}$] Set of branches or transmission lines
    \item[$\mathcal{G}$] Set of generators
\end{abbrv}

\begin{abbrv}
    \item[$\mathbf{j}$] Imaginary unit, $\mathbf{j} = \sqrt{-1}$
    \item[$s^{\text{d}}_{i}$] Complex power demand at bus $i$, $s^{\text{d}}_{i} = p^{\text{d}}_{i} + \mathbf{j} q^{\text{d}}_{i}$
    \item[$Y^{S}_{i}$] Shunt admittance at bus $i$
    \item[$Y_{jk}, Y_{kj}$] Series admittance of branch $(j,k)$
    \item[$Y^{c}_{jk}$] Shunt admittance of branch $(j,k)$
    \item[$\overline{S}_{jk}$] Thermal limit of branch $(j,k)$
    \item[$\underline{s}^{\text{g}}_{i}, \overline{s}^{\text{g}}_{i}$] Lower and upper bounds of complex power generation at bus $i$
    \item[$\underline{V}_{i}, \overline{V}_{i}$] Voltage magnitude bounds at bus $i$
    \item[$\mathbf{z}_{jk}$] Impedance of branch $(j,k)$, $\mathbf{z}_{jk} = r_{jk} + \mathbf{j} x_{jk}$
    \item[$\overline{\theta}_{jk}, \underline{\theta}_{jk}$] Phase angle difference limits for branch $(j,k)$
    \item[$\theta^{M}$] Big-M constant used in phase angle constraints
    \item[$c_{i}$] Generation cost coefficient at bus $i$
    \item[$\mathbf{A}$] Branch-bus incidence matrix
\end{abbrv}

\begin{abbrv}
    \item[$s^{\text{g}}_{i}$] Complex power generation at bus $i$, $s^{\text{g}}_{i} = p^{\text{g}}_{i} + \mathbf{j} q^{\text{g}}_{i}$
    \item[$S_{jk}, S_{kj}$] Complex power flow from bus $j$ to bus $k$ and from bus $k$ to bus $j$
    \item[$V_{i}$] Voltage phasor at bus $i$, $V_{i} = |V_{i}| e^{\mathbf{j} \theta_{i}}$
    \item[$u_{jk}$] Binary variable indicating the status of branch $(j,k)$; $u_{jk} = 1$ if the branch is in service, $u_{jk} = 0$ if it is switched out
    \item[$p_{jk}$] Real power flow on branch $(j,k)$
    \item [$\mathbf{b}$] DC power flow coefficient parameter vector
    \item [$\boldsymbol{\gamma}, \boldsymbol{\rho}, \boldsymbol{\psi}$] DC power flow bias parameter vectors
\end{abbrv}

%% file: text/intro3.tex
\section{Introduction}
 \label{sec:Introduction}

First introduced in the early 1980s~\cite{koglin1981first, koglin1982corrective}, transmission switching has emerged as an important control mechanism for power systems. Intentionally opening transmission lines may appear counterintuitive, but is more generally related to a phenomenon known as Braess's paradox whereby removing edges from a network can improve performance~\cite{cohen1991paradoxical, witthaut2012braess}. 
Building upon this concept, Optimal Transmission Switching (OTS) problems select transmission line statuses to reduce operational costs, manage congestion, and improve voltage stability \cite{o2005dispatchable, fisher2008optimal}.
Transmission switching can mitigate or avoid the need for more costly control mechanisms such as generation rescheduling and load shedding~\cite{Shahidehpour2002market, Conejo2006decomposition}, with applications including managing transmission flow violations~\cite{Bacher1986network}, reducing congestion \cite{granelli2006optimal}, and enhancing system security during contingencies \cite{Schnyder1988integrated, Rolim1999corrective}.
Moreover, OTS can provide significant improvements to economic efficiency~\cite{o2005dispatchable,fisher2008optimal, hedman2008optimal}. When incorporated into expansion planning, OTS can also defer investments and optimize resource utilization~\cite{Khodaei2010}.

Achieving these benefits can be challenging due to the computational difficulties associated with solving OTS problems. The AC-OTS problem minimizes operational costs while adhering to the nonlinear and nonconvex AC power flow equations, voltage limits, and generation capacities, resulting in a computationally challenging Mixed-Integer Nonlinear Program (MINLP)~\cite{capitanescu2011state}.
To address these computational challenges, researchers often employ the DC power flow approximation, which linearizes the AC power flow equations by neglecting reactive power injections, voltage magnitude variations, and trigonometric nonlinearities~\cite{stott2009dc}. This leads to the DC Optimal Transmission Switching (DC-OTS) formulation, expressed as a Mixed-Integer Linear Program (MILP)\textcolor{black}{~\cite{lehmann2014complexity, ruiz2011fast, ruiz2012tractable}}.
\textcolor{black}{In \cite{kocuk2016cycle}, the authors developed a cycle-based formulation with a set of valid inequalities for the DC-OTS problem. Although this approach enables rapid problem-solving, the optimal network configuration from DC transmission switching may not satisfy AC feasibility criteria. Moreover, it may exaggerate cost savings and overlook critical stability issues \cite{hijazi2017convex}. The authors of~\cite{fattahi2018bound} develop a mixed-integer linear programming formulation for DC-OTS. They use a big-M method to reformulate the nonlinear constraints between switchable line power flows and voltage angle differences, demonstrating that identifying the strongest big-M values is NP-hard and proposing an approach for setting them. In \cite{dey2022node}, a set of cutting planes is introduced specifically for the DC-OTS problem.}


While DC-OTS is computationally more tractable, the DC power flow model's inaccuracies can result in solutions that perform poorly (i.e., have higher cost or cause infeasibility) when evaluated using an AC power flow model~\cite{barrows2014,coffrin2014switching}. 
%
%
\textcolor{black}{To address these shortcomings and achieve greater accuracy in capturing AC power flow behavior, researchers have turned to convex relaxation techniques developed for the AC-OPF problem, adapting them to the AC-OTS context. The literature on AC-OTS convex relaxations, though relatively sparse, builds upon the well-established foundation of AC-OPF relaxations. These include the second-order cone (SOC) relaxation~\cite{jabr2006radial}, the quadratic convex (QC) relaxation~\cite{coffrin2015qc}, and the semidefinite programming (SDP) relaxation~\cite{bai2008semidefinite, Lavaei2012, molzahn2013implementation}. For the AC-OPF problem, the SDP and QC relaxations are known to provide stronger (i.e., tighter) objective value bounds than the SOC relaxation, though the relative strength of SDP versus QC relaxations depends on the specific problem instance, as their theoretical guarantees are not directly comparable~\cite{coffrin2015qc}. Computationally, however, the SOC and QC relaxations are generally faster and more reliable than the SDP relaxation, which can suffer from scalability issues due to its higher dimensionality~\cite{coffrin2015qc}. In the AC-OTS setting, where on/off switching decisions introduce binary variables, these relaxations must be extended to mixed-integer formulations, which introduces further complications. To this end, several advanced approaches have been proposed. For instance, the Linear Programming AC (LPAC) approximation offers a computationally efficient alternative that captures some aspects of AC power flow, such as voltage magnitudes, while remaining linear~\cite{Carleton2014}. Similarly, the QC relaxation has been adapted to incorporate on/off decision variables, providing a lower bound on the generation-cost minimization objective in AC-OTS~\cite{coffrin2015qc, hijazi2017convex}. This mixed-integer QC relaxation has been further refined with stronger linearizations and novel valid inequalities, enhancing its ability to approximate the true AC-OTS solution~\cite{guo2022tightening, hijazi2017convex}. The mixed-integer second-order cone programming (MISOCP) relaxation from~\cite{kocuk2017new} also provides a promising approach that balances computational efficiency and solution quality.}

\textcolor{black}{Despite these advances, a trade-off persists: while these alternative power flow models improve upon the DC approximation by better capturing AC nonlinearities, they increase computational complexity and still fall short of fully representing the AC system's behavior. As a result, solutions derived from these relaxations may remain suboptimal or infeasible when evaluated under full AC power flow constraints, particularly in the OTS context where topology changes amplify the impacts of modeling inaccuracies. Ongoing research continues to explore ways to tighten these relaxations and enhance their practical applicability, building on the rich theoretical and computational insights from AC-OPF studies~\cite{bestuzheva2020}.}

In our prior work, we developed an algorithm for improving the accuracy of the DC power flow approximation to better match AC power flow results \cite{taheri2023optimizing}. We extended this algorithm to the DC Optimal Power Flow (DC-OPF) problem by solving a bilevel optimization problem to select parameters that minimize discrepancies in generator setpoints between DC-OPF and AC-OPF solutions~\cite{taheri2024DCOPF}. Using differential optimization techniques inspired by methods for training machine learning models, the algorithm in~\cite{taheri2024DCOPF} significantly improved the accuracy of DC models while maintaining computational efficiency. 
In the context of transmission switching, a natural approach would be to directly apply the optimized parameters from our prior work in~\cite{taheri2023optimizing,taheri2024DCOPF} to the OTS problem, aiming to achieve better results than traditional DC power flow approximations. However, this approach did not yield the expected improvements and, in some cases, performed worse than traditional DC power flow parameter choices. Our intuition is that optimizing parameters to match generator setpoints may not align well with the characteristics of OTS problems, where network congestion and line flows are more critical components for decision making. Essentially, the DC power flow parameters optimized for DC-OPF may be overfitted for purposes not directly relevant to OTS decisions, leading to suboptimal switching actions.

To address this issue, this paper develops a new algorithm for optimizing DC power flow parameters tailored specifically for the OTS problem. This approach leverages the methodology from our previous work in~\cite{taheri2024DCOPF} but introduces key modifications to suit the challenges of OTS. One significant challenge is that computing the optimal AC-OTS solution is computationally infeasible for large systems, making it impractical to set up a loss function that minimizes the error with respect to that solution. Additionally, the OTS problem includes discrete variables, so we cannot directly apply concepts from differentiable optimization as in our previous work.

To overcome these challenges, we focus on optimizing DC power flow parameters for a single loading condition with all computations done online (as opposed to the offline parameter optimization approach in~\cite{taheri2023optimizing,taheri2024DCOPF}). We also develop a proxy for the accuracy of the DC-OTS solution relative to the AC-OTS solution by exploiting the observation that congestion patterns with the nominal topology are major drivers of the lines where switching is influential. Specifically, we define a loss function that aligns DC-OPF line flows with those from the AC-OPF solution, emphasizing congested lines. By optimizing the DC power flow parameters with respect to this loss function, we improve the quality of the discrete switching decisions made by the DC-OTS problem. Furthermore, we introduce and optimize a bias parameter that implicitly accounts for the reactive component of apparent power flows in the AC power flow model. This parameter enhances the DC model’s ability to capture the AC system’s line congestion behavior.

In addition to parameter optimization, we use an approach for implicitly limiting the number of switching decisions.  Rather than fixing a maximum number of line switching operations, we allow switching decisions that the DC-OTS problem predicts will reduce operating costs by a user-defined threshold. This approach permits any number of switching decisions that are expected to have a significant impact while avoiding those expected to yield minimal benefits, enhancing both solution quality and computational efficiency. Using these methodologies, we show that our optimized DC-OTS problem provides solutions that are both lower in cost and more often feasible when evaluated in the AC power flow context, compared to traditional DC-OTS formulations. Our approach also outperforms alternative benchmarks that use more complex power flow approximations and relaxations, such as LPAC and QC models, in terms of both solution quality and computational speed.

The key contributions of this paper are:

\begin{itemize}

\item \textbf{Tailored Parameter Optimization for OTS:} We develop an approach to optimize DC power flow parameters specifically for the OTS problem, enhancing the performance of switching decisions. This approach includes:
\begin{itemize}
    \item \textit{OTS-Focused Loss Function:} We design a loss function that focuses on line flows, making the parameter optimization process sensitive to line congestion which is critical for transmission switching decisions.

    \item \textit{Bias Parameters for Congestion:} We introduce and optimize a new bias parameter to account for the reactive power component in line flows, improving the DC model's representation of apparent power flows.
\end{itemize}

\item \textbf{Implicit Limit on Switching Decisions:} We incorporate an implicit limit on the number of transmission lines that can be switched via a modified cost function. This improves computational efficiency and reflects practical operational requirements.

\item \textbf{Numerical Validation:} We provide numerical results demonstrating that our optimized DC-OTS approach achieves superior accuracy and lower costs compared to traditional DC-OTS, LPAC-OTS, and QC-OTS methods.

\end{itemize}

The rest of the paper is structured as follows. Section~\ref{sec:OPF} reviews the AC-OTS and DC-OTC problems. Section~\ref{sec:Proposed_Algorithm} presents our parameter optimization algorithm for DC-OTS problems. Section~\ref{sec:Numerical experiments} provides numerical results demonstrating the algorithm's performance. Section~\ref{sec:conclusion} concludes the paper.

%% file: text/opf.tex
\section{AC and DC Optimal Transmission Switching}
\label{sec:OPF}

To establish the foundational concepts and notation for the OPF and OTS problems, we begin by defining the mathematical symbols used throughout this section. The real and imaginary parts of a complex number are denoted by $\Re(\cdot)$ and $\Im(\cdot)$, respectively. The complex conjugate is indicated by $(\cdot)^\star$, and the transpose of a matrix is represented by $(\cdot)^\top$. The argument, or phase angle, of a complex number is expressed as $\angle(\cdot)$. Variables with upper and lower bounds are signified using overlines $(\overline{\cdot})$ and underlines $(\underline{\cdot})$.
The power network under consideration consists of buses, transmission lines, and generators which are represented by the sets $\mathcal{N}$, $\mathcal{E}$, and $\mathcal{G}$, respectively. Each bus $i \in \mathcal{N}$ is associated with a voltage phasor $V_i$ and a corresponding phase angle $\theta_i$. The bus characteristics include a shunt admittance $Y_i^S$, a complex power demand $s_i^\text{d} = p_i^\text{d} + \mathbf{j} q_i^\text{d}$, and, when applicable, a generated complex power $s_i^\text{g} = p_i^\text{g} + \mathbf{j} q_i^\text{g}$. For buses without generation capabilities, the generation limits are set to zero.
Transmission lines connecting the buses are denoted by edges $(j, k) \in \mathcal{E}$. The power flows on these lines are represented by $S_{jk}$ and $S_{kj}$. Each line is characterized by its resistance $r_{jk}$ and reactance $x_{jk}$ as well as the associated series admittances $Y_{jk}$ and $Y_{kj}$, along with a shunt admittance $Y_{jk}^c$. In the context of the OTS problem, we introduce a binary variable $u_{jk}$ for each transmission line $(j, k) \in \mathcal{E}$, where $u_{jk} = 1$ indicates that the line is in service and $u_{jk} = 0$ means it is switched out. This variable allows for the optimization of the network topology by enabling or disabling lines to improve system performance.

With the notation and network components defined, we next present the AC and DC formulations of the OPF problem. The AC-OPF problem incorporates the full set of nonlinear AC power flow equations, providing a detailed and accurate model of the network. In contrast, the DC-OPF uses a linearized approximation of these equations which enhances computational efficiency but reduces modeling accuracy.

\subsection{AC-OPF Formulation}
\label{sec:OPF:AC}

The AC-OPF problem is presented in Model~\ref{model:ACOPF}. The objective function in~\eqref{eq:ACOPF:objective} minimizes the total generation cost. Voltage magnitude limits are enforced by constraints~\eqref{eq:ACOPF:voltage_bounds}, and generator output limits are specified in~\eqref{eq:ACOPF:dispatch_bounds}. Thermal limits for power flows are imposed by~\eqref{eq:ACOPF:thermal_limits}. Power balance at each bus is ensured by the nodal power balance equations~\eqref{eq:ACOPF:kirchhoff}, and power flows on each line are governed by the branch power flow equations~\eqref{eq:ACOPF:ohm_fr} and~\eqref{eq:ACOPF:ohm_to}. The reference angle at the slack bus is fixed by~\eqref{eq:ACOPF:ref}, and phase angle differences are limited by~\eqref{eq:ACOPF:phase_angle_limits}. Due to the nonlinearity of the AC power flow equations, the AC-OPF problem is non-convex and typically requires nonlinear optimization techniques such as interior-point methods to find high-quality solutions.

\begin{model}[!t]
    \caption{AC-OPF Problem}
    \label{model:ACOPF}
    \begin{subequations}
    \label{eq:ACOPF}
    \normalsize
    \begin{align}
        \min_{s^{\text{g}}_{i}, V_{i}} \quad 
        & \sum_{i \in \mathcal{N}} c_{2i} \left(\Re(s^{\text{g}}_{i})\right)^{2} + c_{1i}\Re(s^{\text{g}}_{i}) + c_{0i} \label{eq:ACOPF:objective}\\
        \text{s.t.} \quad  &(\forall i\in\mathcal{N}, ~\forall (j,k)\in\mathcal{E}) \nonumber\\
            & \underline{V}_{i} \leq |V_{i}| \leq \overline{V}_{i} 
            &&  \label{eq:ACOPF:voltage_bounds}\\
            & \underline{s}^{\text{g}}_{i} \leq s^{\text{g}}_{i} \leq \overline{s}^{\text{g}}_{i}
            &&  \label{eq:ACOPF:dispatch_bounds}\\
            & \left|S_{jk}\right| \leq \overline{S}_{jk}, \quad \left|S_{kj}\right| \leq \overline{S}_{jk}
            &&  \label{eq:ACOPF:thermal_limits}\\
            & s^{\text{g}}_{i} - s^{\text{d}}_{i} - (Y^{S}_{i})^{\star} |V_{i}|^{2} = \sum_{(i, j) \in \mathcal{E}} S_{ij} + \sum_{(k, i) \in \mathcal{E}} S_{ik}
            &&  \label{eq:ACOPF:kirchhoff}\\
            & S_{jk} = \left(Y_{jk} + Y_{jk}^{c}\right)^{\star} V_{j} V_{j}^{\star} - Y_{jk}^{\star} V_{j} V_{k}^{\star}
            &&  \label{eq:ACOPF:ohm_fr}\\
            & S_{kj} = \left(Y_{jk} + Y_{jk}^{c}\right)^{\star} V_{k} V_{k}^{\star} - Y_{jk}^{\star} V_{j} V_{k}^{\star}
            &&  \label{eq:ACOPF:ohm_to}\\    
            & \theta_{\text{ref}}=0 &&  \label{eq:ACOPF:ref}\\  
            & \underline{\theta}_{jk} \leq \angle \left(V_{j} V_{k}^{\star}\right) \leq \overline{\theta}_{jk}  &&  \label{eq:ACOPF:phase_angle_limits}\\    
            \textbf{Variables}: & \; s^{\text{g}}_{i} \; (\forall i \in \mathcal{N}), \quad V_{i} \; (\forall i \in \mathcal{N})  \nonumber
    \end{align}
    \end{subequations}
\end{model}

\subsection{AC-OTS Formulation}
\label{sec:OTS}

The AC Optimal Transmission Switching (AC-OTS) problem extends the AC-OPF model (Model~\ref{model:ACOPF}) by introducing the ability to switch lines on or off to optimize system performance. This is achieved by adding binary variables $u_{jk}$ for each line $(j,k) \in \mathcal{E}$, where $u_{jk} = 1$ indicates the line is in service and $u_{jk} = 0$ means it is switched out. Incorporating these binary variables transforms the problem into an MINLP.

Most of the constraints in the AC-OTS model presented in Model~\ref{model:ACOTS} are similar to those in the AC-OPF model. The key differences involve the inclusion of the binary variables and the modification of certain constraints to account for line switching.
%
%
The objective~\eqref{eq:ACOPF:objective}, voltage magnitude limits, generator output limits, nodal power balance, and the reference angle constraints remain the same as in the AC-OPF model.
The thermal limits for power flows are modified in constraints~\eqref{eq:ACOTS:thermal_limits} to include the line statuses via the binary variables $u_{jk}$. When a line is switched out ($u_{jk} = 0$), these constraints effectively set the flow limits to zero. 
%
The power flow equations are adjusted in constraints~\eqref{eq:ACOTS:ohm_fr} and~\eqref{eq:ACOTS:ohm_to} to include the binary variables $u_{jk}$, effectively setting the power flows to zero when a line is switched out. 
%
The phase angle difference constraints are modified using a \mbox{big-M} formulation in constraints~\eqref{eq:ACOTS:phase_angle_limits1} and~\eqref{eq:ACOTS:phase_angle_limits2}. These constraints ensure that when a line is in service ($u_{jk} = 1$), the phase angle difference between the connected buses is within the specified limits $\underline{\theta}_{jk}$ and $\overline{\theta}_{jk}$. When a line is switched out ($u_{jk} = 0$), the big-M constant $\theta^{M}$ relaxes the constraints so they do not restrict the phase angle difference. Constraint~\eqref{eq:ACOTS:binary} enforces the binary nature of the switching variables.

\begin{model}[!t]
    \caption{AC-OTS Problem}
    \label{model:ACOTS}
    \begin{subequations}
    \label{eq:ACOTS}
    \normalsize
    \begin{align}
        \min_{s^{\text{g}}_{i}, V_{i}, u_{jk}} \eqref{eq:ACOPF:objective} \label{eq:ACOTS:objective}\\
        \text{s.t.} \quad  &(\forall i\in\mathcal{N}, \quad \forall (j,k)\in\mathcal{E}) \nonumber\\
            & \eqref{eq:ACOPF:voltage_bounds}, \eqref{eq:ACOPF:dispatch_bounds}, \eqref{eq:ACOPF:kirchhoff}, \eqref{eq:ACOPF:ref}\\
            & \left|S_{jk}\right| \leq \overline{S}_{jk} u_{jk}, \quad \left|S_{kj}\right| \leq \overline{S}_{jk} u_{jk}
            &&  \label{eq:ACOTS:thermal_limits}\\
            & S_{jk} = u_{jk} \left[ \left(Y_{jk} + Y_{jk}^{c}\right)^{\star} V_{j} V_{j}^{\star} - Y_{jk}^{\star} V_{j} V_{k}^{\star} \right]\hspace{-3em}
            &&  \label{eq:ACOTS:ohm_fr}\\
            & S_{kj} = u_{jk} \left[ \left(Y_{kj} + Y_{kj}^{c}\right)^{\star} V_{k} V_{k}^{\star} - Y_{kj}^{\star} V_{k} V_{j}^{\star} \right]\hspace{-3em}
            &&  \label{eq:ACOTS:ohm_to}\\      
            & \underline{\theta}_{jk} u_{jk} - \theta^{M}(1-u_{jk}) \leq \angle \left(V_{j} V_{k}^{\star}\right) &&  \label{eq:ACOTS:phase_angle_limits1}\\ 
            & \angle \left(V_{j} V_{k}^{\star}\right) \leq \overline{\theta}_{jk} u_{jk} + \theta^{M}(1-u_{jk}) &&  \label{eq:ACOTS:phase_angle_limits2}\\ 
            & u_{jk} \in \{0,1\} && \label{eq:ACOTS:binary}\\
            \textbf{Variables}: &~ s^{\text{g}}_{i} \; (\forall i \in \mathcal{N}), \quad V_{i}\;  (\forall i \in \mathcal{N}),\quad  u_{jk}\;  (\forall (j,k) \in \mathcal{E}) \nonumber
    \end{align}
    \end{subequations}
\end{model}

The AC-OTS problem is a MINLP due to the binary variables and nonlinear power flow equations, making it more computationally challenging than the AC-OPF problem.
To tackle this computational burden, the problem is often simplified to an MILP by replacing the underlying AC-OPF formulation with the DC-OPF formulation. Next we will discuss the DC-OPF and DC-OTS problems.

\subsection{DC-OPF Formulation}
\label{subsec:OPF:DC}

The DC-OPF model simplifies the AC-OPF problem by linearizing the power flow equations under assumptions of constant voltage magnitudes, small phase angle differences, negligible line losses, and the absence of reactive power \cite{stott2009dc}. Due to its computational efficiency, the DC-OPF is widely used in transmission system analyses, electricity market studies, and operational planning.

\begin{model}[!b]
    \caption{DC-OPF Problem}
    \label{model:DCOPF}
    \begin{subequations}
    \label{eq:DCOPF}
    \normalsize
    \begin{align}
       \min_{p^{\text{g}}_{i}, \theta_{i}} \quad 
        & \sum_{i \in \mathcal{N}} c_{2i} \left(p^{\text{g}}_{i}\right)^{2} + c_{1i} p^{\text{g}}_{i} + c_{0i}  \label{eq:DCOPF:obj}\\
        \text{s.t.} \quad &(\forall i\in\mathcal{N}, \quad \forall (j,k)\in\mathcal{E}) \nonumber\\
        & \underline{p}^{\text{g}}_{i} \leq p^{\text{g}}_{i} \leq \overline{p}^{\text{g}}_{i} 
               && \label{eq:DCOPF:bounds:pg}\\
        & \left|p_{jk}\right| \leq \overline{S}_{jk}
               && \label{eq:DCOPF:bounds:pf}\\       
        & p^{\text{g}}_{i} - p^{\text{d}}_{i} - \gamma_{i} = \sum_{(i,j) \in \mathcal{E}} p_{ij} + \sum_{(k,i) \in \mathcal{E}} p_{ik}   
               && \label{eq:DCOPF:power_balance}\\
        & p_{jk} = b_{jk} (\theta_{k} - \theta_{j}) + \rho_{jk}
               && \label{eq:DCOPF:ohm}\\
        & \underline{\theta}_{jk} \leq \theta_{jk} \leq \overline{\theta}_{jk}  &&  \label{eq:DCOPF:phase_angle_limits}\\  
        & \theta_{\text{ref}} = 0 && \label{eq:DCOPF:ref}\\
        \textbf{Variables}: &~ p^{\text{g}}_{i} \; (\forall i \in \mathcal{N}), \quad \theta_{i} \; (\forall i \in \mathcal{N})  \nonumber
    \end{align}
    \end{subequations}
\end{model}

\begin{model}[!b]
    \caption{DC-OPF in Matrix Form}
    \label{model:DCOPF_matrixformat}
    \begin{subequations}
    \label{eq:DCOPF_matrixformat}
    \normalsize
    \begin{align}
       \min_{\mathbf{p}^{\text{g}}, \boldsymbol{\theta}} \quad 
        & {\mathbf{p}^{\text{g}}}^{\top} \text{diag}(\mathbf{c}_{2}) \mathbf{p}^{\text{g}} + \mathbf{c}_{1}^{\top} \mathbf{p}^{\text{g}} + \sum_{i \in \mathcal{N}} c_{0i}  \label{eq:DCOPF:obj_matrixformat}\\
        \text{s.t.} \quad \nonumber\\
         & \underline{\mathbf{p}}^{\text{g}} \leq \mathbf{p}^{\text{g}} \leq \overline{\mathbf{p}}^{\text{g}}
               && \label{eq:DCOPF:bounds:pg_matrixformat}\\
        & \mathbf{p}^{\text{g}} - \mathbf{p}^{\text{d}} - \boldsymbol{\gamma} = \mathbf{A}^{\top} \left( \text{diag}(\mathbf{b}) \mathbf{A} \boldsymbol{\theta} + \boldsymbol{\rho} \right)  
               && \label{eq:DCOPF:power_balance_matrixformat}\\
        & \left| \text{diag}(\mathbf{b}) \mathbf{A} \boldsymbol{\theta} + \boldsymbol{\rho} \right| \leq \overline{\mathbf{S}}
                && \label{eq:DCOPF:bounds:pf_matrixformat}\\ 
    & \underline{\boldsymbol{\theta}}^{\Delta} \leq \mathbf{A} \boldsymbol{\theta} \leq \overline{\boldsymbol{\theta}}^{\Delta}
           \label{eq:DCOPF:angle_diff}\\
        & \theta_{\text{ref}} = 0
                && \label{eq:DCOPF:theta-ref_matrixformat}\\
        \textbf{Variables}: & \; \mathbf{p}^{\text{g}}, \; \boldsymbol{\theta}  \nonumber
    \end{align}
    \end{subequations}
\end{model}

Model~\ref{model:DCOPF} presents the DC-OPF formulation, where the objective function~\eqref{eq:DCOPF:obj} minimizes total generation costs. Constraints~\eqref{eq:DCOPF:bounds:pg} enforce generator output limits, and constraints~\eqref{eq:DCOPF:bounds:pf} impose thermal limits on transmission lines. The power balance at each bus is maintained by~\eqref{eq:DCOPF:power_balance}, while the linearized power flow equations are given by~\eqref{eq:DCOPF:ohm}. The voltage angle at the reference bus is fixed by~\eqref{eq:DCOPF:ref}.

Model~\ref{model:DCOPF_matrixformat} shows the DC-OPF problem in matrix form. The parameters $\mathbf{b}$, $\boldsymbol{\gamma}$, and $\boldsymbol{\rho}$ influence the accuracy of the DC-OPF approximation. Two common methods for setting these parameters are the cold-start and hot-start approaches.

\subsubsection{Cold-start DC power flow}

In the cold-start approach, parameters are chosen independently of any nominal operating point. The line susceptance $b_{jk}$ is often set as the imaginary part of the negative inverse of the line impedance:
\begin{equation}
\label{eq:cold-start}
b^{\text{cold}}_{jk} = \Im\left(\frac{-1}{r_{jk} + \mathbf{j} x_{jk}}\right).
\end{equation}
Bias terms $\boldsymbol{\gamma}$ and $\boldsymbol{\rho}$ are typically initialized to zero, simplifying the model but potentially reducing accuracy when detailed system information is available.

\subsubsection{Hot-start DC power flow}

The hot-start approach improves the underlying DC power flow model by incorporating data from a nominal AC power flow solution. Parameters are adjusted based on this solution to enhance accuracy. For example, the localized loss modeling approach in \cite{stott2009dc} defines the parameters as:
\begin{subequations}
    \begin{align}
        b^{\text{hot}}_{jk} &= b_{jk} v^{\bullet}_{j} v^{\bullet}_{k} \frac{\sin(\theta^{\bullet}_{j} - \theta^{\bullet}_{k})}{\theta^{\bullet}_{j} - \theta^{\bullet}_{k}}, \label{eq:sub1-hot-start} \\
        \gamma^{\text{hot}}_{j} &= \sum_{(j,k) \in \mathcal{E}} \Re(Y_{jk}) v^{\bullet}_{j} \left(v^{\bullet}_{j} - v^{\bullet}_{k} \cos(\theta^{\bullet}_{j} - \theta^{\bullet}_{k})\right), \label{eq:sub2-hot-start}\\
        \rho^{\text{hot}}_{jk} &= \Re(Y_{jk}) v^{\bullet}_{j} \left(v^{\bullet}_{j} - v^{\bullet}_{k} \cos(\theta^{\bullet}_{j} - \theta^{\bullet}_{k})\right), \label{eq:sub3-hot-start}
    \end{align}%
    \label{eq:hot-start}%
\end{subequations}%
where the superscript $(\,\cdot\,)^{\bullet}$ denotes quantities obtained from the nominal AC power flow solution, \textcolor{black}{and the value of \( b_{ij} \) used on the right-hand side of~\eqref{eq:sub1-hot-start} is typically the cold-start version}. This method provides a more accurate DC power flow model by incorporating operational data.

\subsubsection{Optimized Parameters for DC Power Flow (DCPF)}
\label{sec:OptimizedParameters}

%

Beyond the traditional cold-start and hot-start approaches, DC power flow parameters can be optimized to enhance model accuracy by minimizing a loss function that quantifies the discrepancy between the DC and AC power flow solutions. In our prior work~\cite{taheri2023optimizing}, we introduced an algorithm to optimize the parameters $\mathbf{b}$, $\boldsymbol{\gamma}$, and $\boldsymbol{\rho}$ by minimizing errors in active power line flows over a specified operating range. This algorithm improves the DC power flow model's accuracy across different operating points without explicitly considering its impact on optimization problems like DC-OPF.

We also proposed a bilevel optimization algorithm~\cite{taheri2024DCOPF} that refines the DC power flow parameters based on a loss function involving generator active power setpoints from the DC-OPF solutions relative to those from the AC-OPF solutions. This algorithm aims to align the DC-OPF solutions more closely with the AC-OPF results by adjusting the DC power flow parameters accordingly.

As we will discuss in Section~\ref{sec:Proposed_Algorithm}, we leverage the bilevel optimization framework from~\cite{taheri2024DCOPF} in combination with a new formulation, optimization method, and loss function to compute DC power flow parameters tailored specifically for DC-OTS problems. By focusing on matching apparent power line flows, our algorithm seeks to improve the accuracy of switching decisions in the DC-OTS problem, bringing them closer to those from the AC-OTS solution.


\subsection{Customized DC-OTS (C-DC-OTS) Formulation}
\label{sec:OPF:DC}

The DC-OTS problem extends the DC-OPF model (Model~\ref{model:DCOPF_matrixformat}) by introducing binary decision variables to represent the on/off statuses of transmission lines. 

Model~\ref{model:DCOPFOTS_matrixformat} presents the C-DC-OTS problem in a matrix form. The decision variables in this model include the active power generation $\mathbf{p}^{\text{g}}$, bus voltage angles $\boldsymbol{\theta}$, and binary line status variables $\mathbf{u} \in \{0,1\}^{|\mathcal{E}|}$. Each element $u_{jk}$ in $\mathbf{u}$ corresponds to a transmission line $(j,k) \in \mathcal{E}$, where $u_{jk} = 1$ indicates the line is in service and $u_{jk} = 0$ signifies it is switched out.

\begin{model}[!t]
    \caption{The C-DC-OTS Problem}
    \label{model:DCOPFOTS_matrixformat}
    \begin{subequations}
    \label{eq:DCOPFOTS_matrixformat}
    \normalsize
    \begin{align}
       \min_{\mathbf{p}^{\text{g}} , \boldsymbol{\theta}, \mathbf{u}} \quad 
        & {\mathbf{p}^{\text{g}}}^{\top}\text{diag}(\mathbf{c}_{2}) \mathbf{p}^{\text{g}}+ \mathbf{c}^{\top}_{1}\mathbf{p}^{\text{g}} + \sum_{i \in \mathcal{N}}c_{0i}  + c^{\text{prof}}\mathbf{1}^{\top}\mathbf{u}\label{eq:DCOPFOTS:obj_matrixformat}\\
        \text{s.t.} \quad \nonumber\\
         & \eqref{eq:DCOPF:bounds:pg_matrixformat}, \eqref{eq:DCOPF:theta-ref_matrixformat}\\
        & \text{diag}(\mathbf{b})\mathbf{A}  \boldsymbol{\theta} 
            +\boldsymbol{\rho}+ (\mathbf{1}- \mathbf{u})\circ \mathbf{M}\geq \boldsymbol{p}^{flow} 
                \label{eq:DCOPFOTS:bounds:pf_matrixformat1}\\ 
        & \text{diag}(\mathbf{b})\mathbf{A} \boldsymbol{\theta}+\boldsymbol{\rho} - (\mathbf{1}- \mathbf{u})\circ \mathbf{M}\leq \boldsymbol{p}^{flow} 
            \label{eq:DCOPFOTS:bounds:pf_matrixformat2}\\ 
        &  \mathbf{p}^{\text{g}} - \textbf{p}^{\text{d}} - \boldsymbol{\gamma} = \mathbf{A}^{\top}\boldsymbol{p}^{flow} 
               \label{eq:DCOPFOTS:power_balance_matrixformat}\\
        &|\boldsymbol{p}^{flow} + \boldsymbol{\psi}| \leq \mathbf{u} \circ \overline{\mathbf{S}} 
                       \label{eq:DCOPFOTS:powerflow_matrixformat}\\
        & \underline{\boldsymbol{\theta}}^{\Delta} \circ \mathbf{u} -\boldsymbol{\theta}^{M}(\mathbf{1}-\mathbf{u})\leq \mathbf{A} \boldsymbol{\theta}  \leq \overline{\boldsymbol{\theta}}^{\Delta}\circ \mathbf{u} + \boldsymbol{\theta}^{M}(\mathbf{1}-\mathbf{u})
           \label{eq:DCOTS:angle_diff}\\
        & \underline{\boldsymbol{\theta}} \leq \boldsymbol{\theta} \leq \overline{\boldsymbol{\theta}}
           \label{eq:DCOPFOTS:theta}\\
        & \mathbf{u} \in \{0,1\}^{|\mathcal{E}|}
                \label{eq:DCOPFOTS:z_binary}
    \end{align}
    \end{subequations}
\end{model}

The objective~\eqref{eq:DCOPFOTS:obj_matrixformat} aims to minimize the total generation cost, similar to the DC-OPF model, but with an added term $c^{\text{prof}} \mathbf{1}^{\top} \mathbf{u}$ that associates a cost of $c^{\text{prof}}$ with each line switching operation. The actual cost required to switch a line is usually neglected as it is much less than typical operational costs. As opposed to a literal cost, we use the term $c^{\text{prof}} \mathbf{1}^{\top} \mathbf{u}$ as a modeling tool to ensure that the algorithm will only switch off a transmission line if doing so results in a reduction of the DC-OTS problem's objective function that outweighs the penalty $c^{\text{prof}}$. This avoids the use of explicit limits on the number of lines that can be switched, enabling any number of switching actions which significantly improve outcomes in the DC-OTS problem while avoiding those that yield marginal improvements to the DC-OPF solution. These marginally improving switching actions may lead to infeasibility or suboptimality when evaluated with an AC power flow model.

%
Constraints~\eqref{eq:DCOPFOTS:bounds:pf_matrixformat1} and~\eqref{eq:DCOPFOTS:bounds:pf_matrixformat2} compute the  \textit{active} power flow on the transmission lines, modified to account for line switching. The term $(\mathbf{1} - \mathbf{u}) \circ \mathbf{M}$ involves the Hadamard (element-wise) product, where $\mathbf{M}$ is a vector of sufficiently large constants (big-M values). This term effectively deactivates the power flow constraints for lines that are switched out ($u_{jk} = 0$) by relaxing the bounds.
The power balance equations are maintained by~\eqref{eq:DCOPFOTS:power_balance_matrixformat}, ensuring that the net \textit{active} power injection at each bus equals the sum of power flows on connected lines, adjusted by the bias term $\boldsymbol{\gamma}$.
Constraint~\eqref{eq:DCOPFOTS:powerflow_matrixformat} enforces thermal limits on the transmission lines, where $\overline{\mathbf{S}}$ is the vector of maximum allowable \textit{apparent} power flows. The term $\mathbf{u} \circ \overline{\mathbf{S}}$ ensures that when a line is switched out ($u_{jk} = 0$), its thermal limit is effectively set to zero, preventing any power flow on that line. The parameter $\boldsymbol{\psi}$ accounts for any offsets or adjustments in the power flows due to ignoring reactive power in DC formulation. We will explain how to choose the $\boldsymbol{\psi}$ parameters in the next section. 
Constraints~\eqref{eq:DCOTS:angle_diff} and \eqref{eq:DCOPFOTS:theta} impose the phase angle difference and voltage angle limits,\footnote{Upper and lower limits in \eqref{eq:DCOPFOTS:theta} are set to $\pm0.6$ radians. This choice enables rapid computations. However, relaxing these constraints could potentially enhance the quality of the solutions obtained~\cite{hedman2008optimal}.}
and constraint~\eqref{eq:DCOPFOTS:z_binary} enforces the binary nature of the line status variables.
The DC-OTS (i.e., Model~\ref{model:DCOPFOTS_matrixformat} without $c^{\text{prof}}\mathbf{1}^{\top}\mathbf{u}$) and C-DC-OTS (i.e., Model~\ref{model:DCOPFOTS_matrixformat}) problems are formulated as mixed-integer linear programming (MILP) problems due to the linear nature of the DC power flow equations and the inclusion of binary variables for line statuses. This formulation makes the DC-OTS computationally more tractable than the AC-OTS problem (i.e., Model~\ref{model:ACOTS}), which involves nonlinear power flow equations and is thus an MINLP.

In the next section, we will discuss how to improve the C-DC-OTS performance (i.e., achieving better switching decision) by improving the underlying DC-OPF formulation through optimizing the $\mathbf{b}$ and $\boldsymbol{\psi}$ parameters. 

%% file: text/proposed_algorithm.tex
\section{Parameter Optimization Strategy}
\label{sec:Proposed_Algorithm}

An intuitive approach to improving the results of the DC-OTS problem would be to follow our previous work in~\cite{taheri2023optimizing} by training the DC-OTS parameters across a spectrum of scenarios, using AC-OTS as the ground truth, similar to machine learning training techniques. However, since solving AC-OTS problems is very challenging, this direct approach is impractical. Instead, we optimize the DC-OPF parameters (ignoring the binary variables) to match the AC-OPF results for each loading condition; unlike our previous work in~\cite{taheri2023optimizing} where we optimized over a spectrum of load scenarios \textit{offline}, here we optimize over a single loading condition \textit{online}.

This section presents a strategy for enhancing the accuracy of the DC-OPF model to obtain DC power flow parameters tailored for use in OTS applications. Focusing on aspects that are most relevant to OTS problems, we obtain DC power flow parameters for which the line congestion in DC-OPF solutions closely aligns with AC-OPF solutions.

\begin{figure}[!t]
\centering
\includegraphics[width=0.48\textwidth]{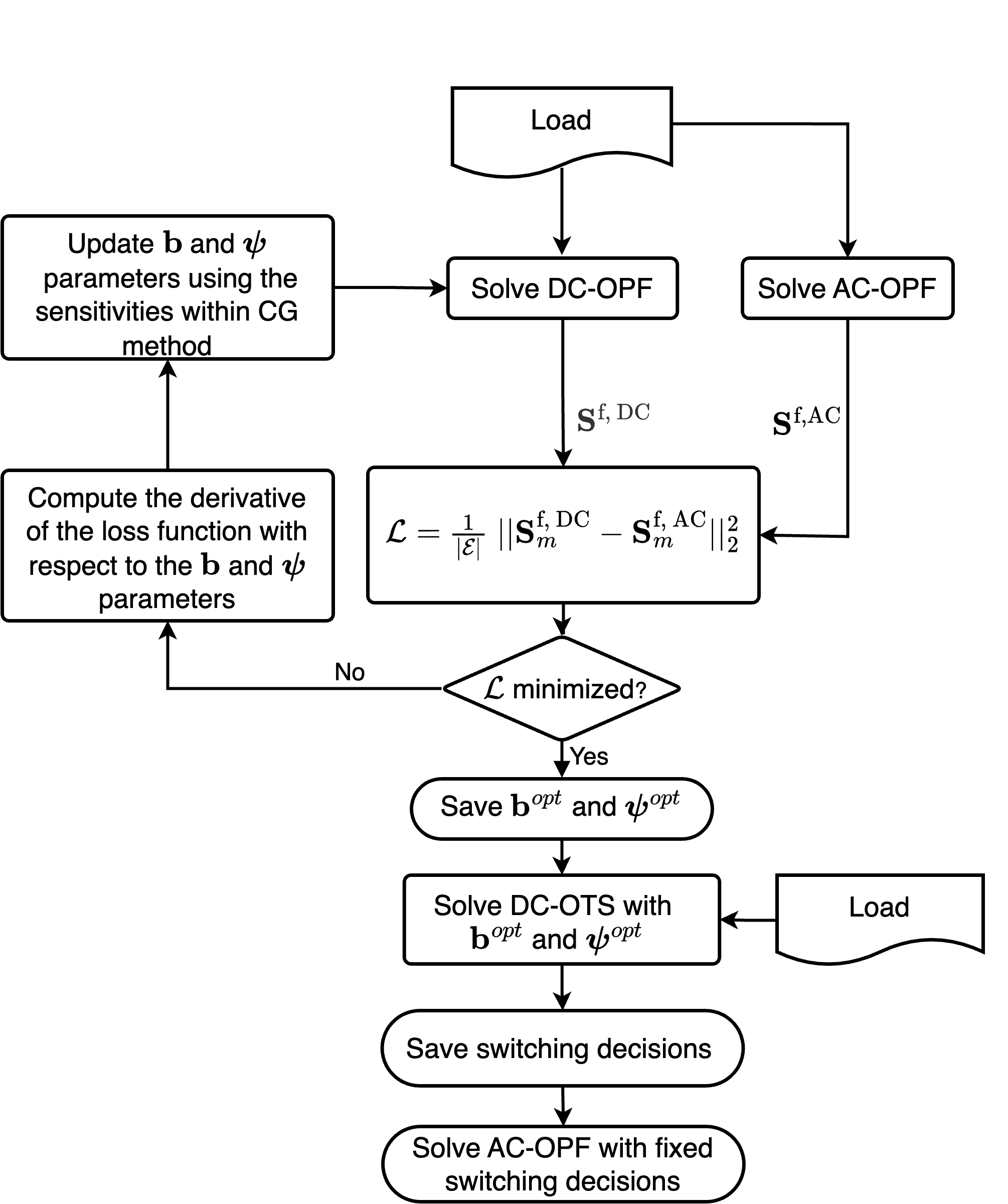}
\caption{Flowchart of the proposed parameter optimization algorithm.}
\label{fig:flowchart}
\end{figure}

The strategy consists of three phases, as illustrated in Fig.~\ref{fig:flowchart}. The first phase optimizes the DC-OPF parameters $\mathbf{b}$ and $\boldsymbol{\psi}$ by minimizing a specially designed loss function that captures discrepancies in the line flows between the DC-OPF and AC-OPF solutions. Including line flows in the loss function makes the model OTS-aware, as line congestion plays a crucial role in transmission switching decisions. In the second phase, the optimized parameters are employed to solve DC-OTS problems efficiently, leveraging the improved accuracy of the DC-OPF model in representing the AC-OPF behavior. In the third phase, the line switching decisions obtained from the DC-OTS solution are applied to an AC-OPF problem with the switched-off lines removed from service. This step ensures that the final solution satisfies the AC power flow equations and adheres to all AC-OPF inequality constraints, providing a feasible and reliable operating point for the power system.

\subsection{OTS-Aware Loss Function Definition}
\label{subsec:Loss Function}

To select DC power flow parameters tailored to the OTS application, we define a loss function that penalizes discrepancies between the apparent power flows in the AC-OPF solution and the real power flows (adjusted by $\boldsymbol{\psi}$) in the DC-OPF solution. Line flows are particularly important in OTS because they determine line congestion, which is a key factor in transmission switching decisions. By basing the loss function on line flow discrepancies, we are encouraging the optimized parameters $\mathbf{b}$ and $\boldsymbol{\psi}$ to capture the critical congestion aspects relevant to OTS. 
\textcolor{black}{We note that the parameters \(\boldsymbol{\gamma}\) and \(\boldsymbol{\rho}\) do not significantly affect DC-OTS performance in this context, in contrast to previous works \cite{taheri2023optimizing, taheri2024DCOPF}. In our approach, we experimented with including these parameters; however, our numerical results indicated that their inclusion often led to overfitting and, in some cases, degraded performance. Consequently, we set \(\boldsymbol{\gamma}\) and \(\boldsymbol{\rho}\) to zero, and focus on optimizing only \(\mathbf{b}\) and \(\boldsymbol{\psi}\).}
The loss function is: 
\begin{align}
\mathcal{L}(\mathbf{b}, \boldsymbol{\psi}) =  \frac{1}{|\mathcal{E}| } \left\| \mathbf{S}^{\text{f,DC}}(\mathbf{b}, \boldsymbol{\psi}) - \mathbf{S}^{\text{f,AC}} \right\|^2_2.
\label{eq:loss_function}
\end{align}

The vector $\mathbf{S}^{\text{f,DC}}(\mathbf{b}, \boldsymbol{\psi})$ represents the approximation of apparent power flows obtained from the DC-OPF solution (i.e., $\boldsymbol{p}^{flow} + \boldsymbol{\psi}$ in~\eqref{eq:DCOPFOTS:powerflow_matrixformat}) and $\mathbf{S}^{\text{f,AC}}$ contains the corresponding values from the AC-OPF solution.
By minimizing this loss function, we aim to adjust the DC-OPF parameters so that the DC-OPF model accurately captures line congestion patterns observed in the AC-OPF solutions. This is crucial for OTS applications, where accurate representation of line flows influences the identification of beneficial switching actions.

The parameter optimization problem is then formulated as:
\begin{equation}
\min_{\mathbf{b}, \boldsymbol{\psi}} \quad \mathcal{L}(\mathbf{b}, \boldsymbol{\psi}).
\label{eq:parameter_optimization_problem}
\end{equation}

Through the sensitivity analysis discussed in the following subsection, we modify the coefficient and bias parameters ($\mathbf{b}$ and $\boldsymbol{\psi}$) to improve the DC-OPF problem’s parameters. As outlined in Model~\ref{model:O-DCOPF_matrixformat}, this process is structured as a bilevel optimization where the upper-level optimizes parameters using the results from the lower-level DC-OPF problems.

\begin{model}[!t]
    \caption{Optimizing DC-OPF Problem}
    \label{model:O-DCOPF_matrixformat}
    \begin{subequations}
    \label{eq:O-DCOPF_matrixformat}
    \normalsize
    \begin{align}
     \min_{\mathbf{b}, \boldsymbol{\psi}} \quad & \mathcal{L} = \frac{1}{|\mathcal{E}|} ||\mathbf{S}^{\text{f},\text{DC}} - \mathbf{S}^{\text{f},\text{AC}}||^2_2 \label{eq:O-DCOPF:main_objective}\\ %
    \text{s.t.} \quad \nonumber \\
        &\min_{\mathbf{p}^{\text{g}} , \boldsymbol{\theta}} \quad  {\mathbf{p}^{\text{g}}}^{\top}\text{diag}(\mathbf{c}_{2}) \mathbf{p}^{\text{g}}+ \mathbf{c}^{\top}_{1}\mathbf{p}^{\text{g}} + \sum_{i \in \mathcal{G}}c_{0i}  \label{eq:O-DCOPF:obj_opt}\\
        &\quad\text{s.t.} \quad \nonumber\\
            & \quad \quad \mathbf{p}^{\text{g}} - \textbf{p}^{\text{d}} = \mathbf{A}^{\top} \Big(\text{diag}(\mathbf{b}) \mathbf{A} \boldsymbol{\theta} \Big)   
               \label{eq:O-DCOPF:power_balance_matrixformat}\\
            & \quad \quad \mathbf{S}^{\text{f},\text{DC}}   =  \text{diag}(\mathbf{b})\mathbf{A} \boldsymbol{\theta} + \boldsymbol{\psi}\label{eq:O-DCOPF:D_DC}\\
            & \quad \quad |\mathbf{S}^{\text{f},\text{DC}}| \leq \bar{\mathbf{S}} 
                \label{eq:O-DCOPF:bounds:pf_matrixformat}\\
            & \quad \quad \underline{\textbf{p}}^{\text{g}} \leq \mathbf{p}^{\text{g}} \leq \overline{\textbf{p}}^{\text{g}}
               \label{eq:O-DCOPF:bounds:pg_matrixformat}\\
               & \quad \quad\underline{\boldsymbol{\theta}^{\Delta}} \leq \mathbf{A} \boldsymbol{\theta} \leq \overline{\boldsymbol{\theta}^{\Delta}}
           \label{eq:O-DCOPF:angle_diff}\\
        & \quad \quad\underline{\boldsymbol{\theta}} \leq \boldsymbol{\theta} \leq \overline{\boldsymbol{\theta}}
           \label{eq:O-DCOPF:theta}\\
               &\quad \quad \theta_{ref} = 0 \label{eq:O-DCOPF:angle_ref}\\
               &\quad  \forall m \in \mathcal{M} \nonumber
    \end{align}
    \end{subequations}
\end{model}

\subsection{Sensitivity Analysis for Optimizing Parameters}
\label{subsec:Sensitivity Analysis}

\textcolor{black}{To solve the optimization problem defined in \eqref{eq:parameter_optimization_problem}, we conduct a sensitivity analysis to compute the gradients of the loss function \(\mathcal{L}\) with respect to the parameters \(\mathbf{b}\) and \(\boldsymbol{\psi}\). These gradients are essential for iteratively updating the parameters using gradient-based optimization techniques to minimize the discrepancy between the DC-OPF and AC-OPF solutions.}

\textcolor{black}{The gradients are given by:}
\begin{subequations}
\color{black}
\begin{align}
\mathbf{g}^{b} = \frac{\partial \mathcal{L}}{\partial \mathbf{b}} = \frac{2}{|\mathcal{E}|} \left( \frac{\partial \mathbf{S}^{\text{f,DC}}}{\partial \mathbf{b}} \left( \mathbf{S}^{\text{f,DC}} - \mathbf{S}^{\text{f,AC}} \right) \right), \label{eq:gradient_b} \\
\mathbf{g}^{\psi} = \frac{\partial \mathcal{L}}{\partial \boldsymbol{\psi}} = \frac{2}{|\mathcal{E}|} \left( \frac{\partial \mathbf{S}^{\text{f,DC}}}{\partial \boldsymbol{\psi}} \left( \mathbf{S}^{\text{f,DC}} - \mathbf{S}^{\text{f,AC}} \right) \right), \label{eq:gradient_gamma}
\end{align}
\end{subequations}
\textcolor{black}{where \(\mathbf{S}^{\text{f,DC}}\) represents the power flows from the DC-OPF solution, \(\mathbf{S}^{\text{f,AC}}\) denotes the corresponding AC-OPF power flows, and \(|\mathcal{E}|\) is the number of edges in the power network.}

\textcolor{black}{The partial derivatives \(\frac{\partial \mathbf{S}^{\text{f,DC}}}{\partial \mathbf{b}}\) and \(\frac{\partial \mathbf{S}^{\text{f,DC}}}{\partial \boldsymbol{\psi}}\) are obtained by differentiating through the DC-OPF optimization problem. We employ differentiable convex optimization techniques, as outlined in foundational works such as \cite{fiacco1990nonlinear, robinson1980strongly, amos2017optnet}, and implemented practically in \cite{cvxpylayers2019}. The remainder of this subsection overviews these techniques.}

\textcolor{black}{
The DC-OPF is a convex quadratic program with variables \(\mathbf{p}^{\text{g}}\) and \(\boldsymbol{\theta}\). The DC-OPF's solution satisfies the Karush-Kuhn-Tucker (KKT) conditions~\cite{karush1939minima, kuhn1951proceedings}, expressed generically as:
\[ f(\mathbf{x}, \boldsymbol{\lambda}; \mathbf{b}, \boldsymbol{\psi}) = 0, \]
where \(\mathbf{x} = [\mathbf{p}^{\text{g}}, \boldsymbol{\theta}]\) and \(\boldsymbol{\lambda}\) denotes the dual variables. The power flows \(\mathbf{S}^{\text{f,DC}} = \text{diag}(\mathbf{b}) \mathbf{A} \boldsymbol{\theta} + \boldsymbol{\psi}\) are computed from the solution. Under regularity conditions such as the Linear Independence Constraint Qualification (LICQ) and strict feasibility, the implicit function theorem ensures that \(\mathbf{x}\) is a differentiable function of \(\mathbf{b}\) and \(\boldsymbol{\psi}\). Differentiating the KKT system with respect to \(\mathbf{b}\) and \(\boldsymbol{\psi}\) yields:
\[ \begin{bmatrix} \frac{\partial f}{\partial \mathbf{x}} & \frac{\partial f}{\partial \boldsymbol{\lambda}} \end{bmatrix} \begin{bmatrix} \frac{\partial \mathbf{x}}{\partial \mathbf{u}} \\ \frac{\partial \boldsymbol{\lambda}}{\partial \mathbf{u}} \end{bmatrix} = -\frac{\partial f}{\partial \mathbf{u}}, \]
where $\mathbf{u} = [\mathbf{b}, \boldsymbol{\psi}]$. Solving this system provides $\frac{\partial \mathbf{x}}{\partial \mathbf{b}}$ and $\frac{\partial \mathbf{x}}{\partial \boldsymbol{\psi}}$, from which the partial derivatives with respect to $\boldsymbol{\theta}$ and the final expressions for $\frac{\partial \mathbf{S}^{\text{f,DC}}}{\partial \mathbf{b}}$ and $\frac{\partial \mathbf{S}^{\text{f,DC}}}{\partial \boldsymbol{\psi}}$ are obtained. In practice, CVXPY’s differentiable convex optimization layers \cite{cvxpylayers2019} canonicalize the DC-OPF into an affine-solver-affine (ASA) form---an affine map to cone program data, a conic solver, and an affine retrieval step---thus computing these derivatives efficiently in a single backward pass.
}

By calculating these gradients, we quantify how perturbations in \(\mathbf{b}\) and \(\boldsymbol{\psi}\) influence the DC-OPF power flows \(\mathbf{S}^{\text{f,DC}}\) and, consequently, the loss function \(\mathcal{L}\). This sensitivity information enables gradient-based optimization algorithms (e.g., stochastic gradient descent) to adjust the parameters in directions that minimize the loss, aligning the DC-OPF solution more closely with the AC-OPF solution.

\subsection{Optimization Algorithm}
\label{subsec:Optimization Algorithm}

To minimize the loss function defined in \eqref{eq:loss_function} and solve Model~\ref{model:O-DCOPF_matrixformat}, Algorithm~\ref{alg:conjugate_gradient} employs the Conjugate Gradient (CG) method~\cite{nocedal2006numerical}.
The CG method iteratively updates the parameters $\mathbf{b}$ and $\boldsymbol{\psi}$ by moving along conjugate directions computed using gradient information while avoiding the computational burden of calculating and storing the Hessian matrix, making it well-suited for large-scale nonlinear optimization.
%

The optimization procedure begins with the initialization of estimates for the parameters $\mathbf{b}$ and $\boldsymbol{\psi}$. Next, in each iteration, the gradients $\mathbf{g}^{b}$ and $\mathbf{g}^{\psi}$ are computed using sensitivity analysis, as described in Section~\ref{subsec:Sensitivity Analysis}. The CG method is then used to update the parameters by moving along conjugate directions determined from the gradient information. After each update, a convergence check is performed by evaluating the loss function to determine if the change is below a predefined threshold. If this criterion is met, the optimization terminates; otherwise, the process repeats. Through this iterative refinement, the parameters are optimized to minimize the loss function, thus improving the accuracy of the DC-OPF model in closely approximating AC-OPF solutions for OTS applications.

\begin{algorithm}[!th]
\caption{Conjugate Gradient (CG) Method}
\label{alg:conjugate_gradient}
\small
\DontPrintSemicolon

\KwInput{
    $\mathbf{x}_0 = [\mathbf{b}^{\top}_{0}, \boldsymbol{\psi}^{\top}_{0}]^{\top}$: Initial guess \\
    \hspace{1cm}$\epsilon$: Tolerance for convergence \\
    \hspace{1cm}$\mathit{max\_iter}$: Maximum iterations \\
    \hspace{1cm}$\mathcal{L}(\mathbf{x})$: Loss function \\
    \hspace{1cm}$\nabla \mathcal{L}(\mathbf{x})$: Gradient function \\
    \hspace{1cm}$\alpha_1$: Armijo condition constant (e.g., $10^{-4}$) \\
    \hspace{1cm}$\alpha_2$: Curvature condition constant, between $\alpha_1$ and $1$
}

\KwOutput{
    Optimized parameters $\mathbf{x}^*$
}

Initialize $\mathbf{x}_k \leftarrow \mathbf{x}_0$, compute $\mathbf{g}_0 \leftarrow \nabla \mathcal{L}(\mathbf{x}_0)$, set $\mathbf{p}_0 \leftarrow -\mathbf{g}_0$, $k \leftarrow 0$\;

\While{$k \leq \mathit{max\_iter}$ \textbf{and} $\|\mathbf{g}_k\| > \epsilon$}{
    \tcp*{Wolfe Line Search to determine $\alpha_k$}
    $\alpha_k \gets 1$\;
    \While{True}{
        \If{$\mathcal{L}(\mathbf{x}_k + \alpha_k \mathbf{p}_k) \leq \mathcal{L}(\mathbf{x}_k) + \alpha_1 \alpha_k \mathbf{g}_k^\top \mathbf{p}_k$ \textbf{and} $\left|\nabla \mathcal{L}(\mathbf{x}_k + \alpha_k \mathbf{p}_k)^\top \mathbf{p}_k\right| \leq \alpha_2 \left|\mathbf{g}_k^\top \mathbf{p}_k\right|$}{
            \textbf{break}\;
        }
        $\alpha_k \gets \alpha_k / 2$\; 
    }
    $\mathbf{x}_{k+1} \gets \mathbf{x}_k + \alpha_k \mathbf{p}_k$\;
    Compute $\mathbf{g}_{k+1} \gets \nabla \mathcal{L}(\mathbf{x}_{k+1})$\;
    \tcp*{Compute Polak-Ribiere coefficient}
    $\beta_k \gets \dfrac{\mathbf{g}_{k+1}^\top (\mathbf{g}_{k+1} - \mathbf{g}_k)}{\mathbf{g}_k^\top \mathbf{g}_k}$\;
    $\mathbf{p}_{k+1} \gets -\mathbf{g}_{k+1} + \beta_k \mathbf{p}_k$\;
    $k \gets k + 1$\;
}

$\mathbf{x}^{*} \leftarrow \mathbf{x}_k$\;

\end{algorithm}


\textcolor{black}{We note that while problem \eqref{eq:loss_function} is nonconvex, initializing the parameters with cold‑start or hot‑start DC heuristics—and observing consistent convergence across these initializations—indicates that our parameter learning algorithm reliably identifies high‑quality solutions, yielding robust improvements in AC feasibility and cost for OTS. Empirical results across various test cases show that our optimized model consistently improves switching decisions and cost outcomes, validating the effectiveness of the optimization process despite the potential for local optima.}

%% file: text/results.tex
\section{Numerical Experiments}
\label{sec:Numerical experiments}

\begin{table*}[!t]\centering
\caption{Opened Lines for Different OTS Methods (\lowercase{``ns.''}: not solved within time limit; \lowercase{``scb.''}: solution cannot be built)}
\smaller
\setlength{\tabcolsep}{2pt}
\renewcommand{\arraystretch}{1.3}
\begin{tabular}{l|m{4.5cm}|m{1.5cm}|m{1.5cm}|m{1.5cm}|m{1.5cm}|m{4.5cm}}

\toprule
\textbf{Case} & \textbf{DC-OTS} & \textbf{LPAC-OTS} & \textbf{QC-OTS} & \textbf{C-DC-OTS} & \textbf{O-DC-OTS} & \textbf{AC-OTS} \\
\midrule
\texttt{3-lmbd-api} & 3 & 3 & -- & 3 & 3 & 3 \\
\hline
\texttt{5-pjm-api} & 3 & ns. & -- & 3 & 3 & 3 \\
\hline
\texttt{14-ieee-api} & 12, 13, 14 & 11, 20 & -- & 14 & -- & -- \\
\hline
\texttt{24-ieee-rts-api} & 2, 9, 14, 30, 34, 35 & 6, 8, 9, 24, 30, 33, 34, 35 & 30, 34, 35 & 2, 9, 14, 30, 34, 35 & 9, 30, 34, 35 & 8, 9, 24, 30, 34, 35 \\
\hline
\texttt{30-as-api} & 8, 11, 22, 32 & 11, 20, 22 & -- & 11, 22 & 11, 22 & 11, 22 \\
\hline
\texttt{30-ieee-api} & 10, 11, 13, 14, 16, 23, 28, 29, 37 & 11, 23, 31 & -- & 12, 26 & 26 & 12, 23, 26 \\
\hline
\texttt{39-epri-api} & 17, 22, 31, 43 & 17, 31 & -- & 6, 16 & 4, 13 & 6, 31 \\
\hline
\texttt{57-ieee-api} & -- & 6, 34, 62, 77 & 13, 32, 52, 73, 77 & -- & -- & 13, 34, 52, 62, 73 \\
\hline
\texttt{60-c-api} & 2, 3, 4, 8, 16, 20, 25, 44, 48, 49, 70, 85 & 3, 4, 17, 20, 25, 44, 86 & 19, 20, 82 & 17, 44 & 16 & 3, 4, 16 \\
\hline
\texttt{73-ieee-rts-api} & \begin{minipage}[t]{\linewidth}1, 3, 6, 14, 24, 27, 34, 35, 36, 37, 38, 50, 65, 73, 74, 75, 76, 85, 88, 98, 107, 111, 113\end{minipage} & ns. & 34, 35, 36, 73, 74 & 36, 37, 55 & 9, 50 & \begin{minipage}[t]{\linewidth}1, 24, 27, 34, 35, 36, 37, 50, 65, 73, 74, 75, 76, 80, 113, 114, 119\end{minipage} \\
\hline
\texttt{89-pegase-api} & \begin{minipage}[t]{\linewidth}
3, 4, 8--10, 12--13, 15--17, 19--21, 24--45, 47--51, 53--54, 63--64, 68, 70, 72--74, 77--78, 87--90, 92--93, 96--103, 106, 108, 111, 113--119, 121--125, 127, 129, 131--136, 138--140, 142, 144--152, 155, 157--158, 161, 165, 167, 169, 172, 175--176, 184--190, 192--193, 197, 199--200
\end{minipage} & scb. & ns. & 9, 34 & 9, 34 & \begin{minipage}[t]{\linewidth}
11, 13, 20, 22, 27, 28, 33, 38--41, 43, 45--46, 53, 76--77, 88--89, 92--93, 99, 101, 107, 109, 111, 116, 118--125, 127--129, 134, 138--139, 141--142, 149, 151, 155, 159, 169, 173, 184, 187--188, 190, 192, 196, 200
\end{minipage} \\
\hline
\texttt{118-ieee-api} & \begin{minipage}[t]{\linewidth}
13, 17--18, 21, 24, 26, 43, 49, 57--58, 72, 75--76, 79--80, 82, 84--86, 89, 91--92, 99, 101--102, 105--106, 111, 117, 126--127, 148--149, 151, 154, 156, 169, 179--180, 186
\end{minipage} & ns. & ns. & 12, 21, 24, 57, 58, 117, 155 & 12, 13, 21, 24, 57, 58, 117, 155 & \begin{minipage}[t]{\linewidth}
1, 12, 14--15, 19, 24, 26, 43--44, 57--58, 65, 69, 72, 79--80, 82, 84--87, 100, 102--103, 106, 117, 127, 148--149, 157, 176, 180, 186
\end{minipage} \\
\hline
\texttt{179-goc-api} & \begin{minipage}[t]{\linewidth}
13, 18, 21--22, 25, 31, 33--36, 45, 47, 49, 53--54, 56, 58, 60--61, 63, 68, 70, 72--73, 75--77, 80, 82, 85--88, 94, 97, 99, 102, 107, 110--111, 114, 118, 125--126, 133, 136, 139, 144, 149--151, 159, 161, 167, 175, 178--179, 185, 189, 191, 197--198, 204--206, 217, 220--221, 223--224, 226, 230--231, 234--237, 240, 252, 255, 259, 261--262
\end{minipage} & scb. & ns. & 159, 160, 161, 186, 202, 204 & 167, 240, 244, 247 & ns. \\
\hline
\texttt{200-activ-api} & 4, 14, 17--18, 24, 38, 50, 58, 62, 70, 76, 86--88, 115--116, 119, 122, 125, 128, 131, 133--134, 138, 153, 162, 164, 182, 190, 204, 219, 221--223, 225--226, 238, 244--245 & ns. & -- & -- & -- & -- \\
\hline
\texttt{240-pserc-api} & \begin{minipage}[t]{\linewidth}
ns.
\end{minipage} & ns. & ns. & 21, 172, 173, 183, 275, 294, 295, 365, 392 & 172, 173, 178, 183, 275, 294, 295& ns. \\
\hline
\texttt{300-ieee-api} & \begin{minipage}[t]{\linewidth}
9, 11--13, 39--40, 42, 51, 54, 64, 69, 72--73, 75--77, 80, 82, 89, 91, 96, 103, 110, 112, 119, 125, 131--132, 138, 146--147, 153, 158, 175, 178--180, 182--184, 186, 189, 192--193, 201, 203, 212, 216, 228, 235--236, 247, 259, 265, 277--281, 283, 288, 299, 302, 305, 313, 320, 324--325, 327, 336, 341, 347--348, 351--352, 357, 368, 375, 377--378, 385, 389, 407
\end{minipage} & ns. & ns. & 190, 197, 199, 203, 204 & -- & \begin{minipage}[t]{\linewidth}
49, 55, 110, 112, 142, 145, 148--149, 151, 163--164, 169--170, 172, 178--179, 197, 199, 201, 203--204, 218, 234, 271, 279, 282, 287, 298, 302--303, 329, 339, 341, 389
\end{minipage} \\
\hline
\texttt{500-goc-api} & ns. & ns. & ns. & 88, 136, 174, 203, 537 & 201 & ns. \\
\hline
\texttt{1354-pegase-api} & ns. & ns. & ns. & 119, 1161, 1562 & 119, 1161, 1562 & ns. \\

\bottomrule
\end{tabular}
\begin{tablenotes}
    \item[*] \scriptsize DC-OTS, LPAC-OTS~\cite{Carleton2014}, QC-OTS~\cite{coffrin2015qc}, and AC-OTS are solved using \texttt{PowerModels.jl}.
\end{tablenotes}
\label{table:opened_lines_methods}
\end{table*}

\begin{table*}[!t]\centering
\caption{Cost and Computation Time Comparison for Different OTS Methods}
\setlength{\tabcolsep}{3pt}
\renewcommand{\arraystretch}{1.2}
\begin{adjustbox}{width=\textwidth}
\begin{threeparttable}
\begin{tabular}{l|c|ccc|cc|c|cccccc}
\toprule
\multirow{2}{*}{\textbf{Case}} & \textbf{AC-OPF} & \multicolumn{6}{c|}{\textbf{AC-OPF with}} & \multicolumn{6}{c}{\textbf{Time (s)}} \\
\cmidrule(lr){3-8} \cmidrule(lr){9-14}
& Cost (\$) & DC-OTS & LPAC-OTS & QC-OTS & C-DC-OTS & O-DC-OTS & AC-OTS & DC-OTS & LPAC-OTS & QC-OTS & C-DC-OTS & \multicolumn{1}{c}{O-DC-OTS} & AC-OTS \\
\midrule
\texttt{3-lmbd-api} & \$11,236 & \$10,636 & \$10,636 & \$11,236 & \$10,636 & \$10,636 & \$10,636 & 0.004 & 0.007 & 0.137 & 0.006 & 0.142~+~0.007 & 0.971 \\
& & (\textbf{--5.34\%}) & (\textbf{--5.34\%}) & (\textbf{0.00\%}) & (\textbf{--5.34\%}) & (\textbf{--5.34\%}) & (\textbf{--5.34\%}) & & & & & &  \\

\texttt{5-pjm-api} & \$76,377 & \$75,190 & ns. & \$76,377 & \$75,190 & \$75,190 & \$75,190 & 0.056 & 1.747 & 0.076 & 0.007 & 2.266~+~0.006 & 0.521 \\
& & (\textbf{--1.55\%}) &  & (\textbf{0.00\%}) & (\textbf{--1.55\%}) & (\textbf{--1.55\%}) & (\textbf{--1.55\%}) & & & & &&  \\

\texttt{14-ieee-api} & \$5,999 & Inf. & Inf. & \$5,999 & Inf. & \$5,999 & \$5,999 & 0.060 & 0.139 & 6.042 & 0.027 & 0.645~+~0.030 & 0.100 \\
& &  &  & (\textbf{0.00\%}) &  & (\textbf{0.00\%}) & (\textbf{0.00\%}) & & & & & & \\

\texttt{24-ieee-rts-api} & \$134,944 & Inf. & \$125,035 & \$124,601 & Inf. & \$122,283 & \$119,743 & 0.151 & 2.482 & 546.8 & 0.106 & 29.96~+~0.107 & 4.446 \\
& &  & (\textbf{--7.34\%}) & (\textbf{--7.66\%}) &  & (\textbf{--9.39\%}) & (\textbf{--11.27\%}) & & & & & & \\

\texttt{30-as-api} & \$4,996 & Inf. & \$2,809 & \$4,996 & \$2,797 & \$2,797 & \$2,797  & 0.073 & 0.939 & 49.03 & 0.083 & 1.836~+~0.080 & 7.141 \\
& &  & (\textbf{--43.77\%}) & (\textbf{0.00\%}) & (\textbf{--44.01\%}) & (\textbf{--44.01\%}) & (\textbf{--44.01\%}) & & & & & & \\

\texttt{30-ieee-api} & \$18,044 & Inf. & Inf. & \$18,044 & \$17,959 & \$17,939 & \$17,936 & 0.059 & 2.195 & 64.65 & 0.088 & 2.100~+~0.090 & 3.230 \\
& &  &  & (\textbf{0.00\%}) & (\textbf{--0.47\%}) & (\textbf{--0.58\%}) & (\textbf{--0.60\%}) & & & & & &  \\

\texttt{39-epri-api} & \$249,672 & Inf. & Inf. & \$249,672 & Inf. & \$246,850 & \$246,723 & 0.028 & 0.759 & 41.59 & 0.046 & 14.61~+~0.045 & 5.234 \\
& &  &  & (\textbf{0.00\%}) &  & (\textbf{--1.13\%}) & (\textbf{--1.18\%}) & & & & & &  \\

\texttt{57-ieee-api} & \$49,290 & \$49,290 & \$49,290 & \$49,279 & \$49,290 & \$49,290 & \$49,274 & 0.031 & 25.98 & 827.2 & 0.021 & 1.849~+~0.023 & 14.17 \\
& & (\textbf{0.00\%}) & (\textbf{0.00\%}) & (\textbf{--0.02\%}) & (\textbf{0.00\%}) & (\textbf{0.00\%}) & (\textbf{--0.03\%}) & & & & & &  \\

\texttt{60-c-api} & \$185,239 & \$183,302 & \$182,145 & \$186,409 & \$182,231 & \$182,028 & \$182,028 & 0.124 & 2.899 & 609.8 & 0.300 & 12.57~+~0.309 & 15.79 \\
& & (\textbf{--1.05\%}) & (\textbf{--1.67\%}) & (\textbf{+0.63\%}) & (\textbf{--1.62\%}) & (\textbf{--1.73\%}) & (\textbf{--1.73\%}) & & & & & &  \\

\texttt{73-ieee-rts-api} & \$422,627 & Inf. & ns. & \$416,549 & Inf. & \$413,133 & \$385,194 & 16.98 & tl. & 21,042 & 4.237 & 18.56~+~4.329 & 611.0 \\
& &  &  & (\textbf{--1.44\%}) &  & (\textbf{--2.25\%}) & (\textbf{--8.86\%}) & & & & & &  \\

\texttt{89-pegase-api} & \$130,175 & Inf. & scb. & ns. &  \$100,702 & \$100,702 & \$100,344 & 5.787 & scb. & tl. & 0.576 & 22.88~+~0.598 & 1,867 \\
& &  &  &  & (\textbf{--22.64\%}) & (\textbf{--22.64\%}) & (\textbf{--22.92\%}) & & & & & &  \\

\texttt{118-ieee-api} & \$242,237 & Inf. & ns. & ns. & \$195,923 & \$195,918 & \$180,312 & 1,720 & tl. & tl. & 18.02 & 84.43~+~17.98 & 3,098 \\
& &  &  &  & (\textbf{--19.12\%}) & (\textbf{--19.12\%}) & (\textbf{--25.56\%}) & & & & & & \\

\texttt{179-goc-api} & \$1,932,044 & Inf. & scb. & ns. & Inf. & \$1,931,004 & ns. & 3.043 & scb. & tl. & 142.3 & 68.44~+~131.4 & tl. \\
& &  &  &  &  & (\textbf{--0.05\%}) &  & & & & & &  \\

\texttt{200-activ-api} & \$35,701 & \$35,945 & ns. & \$35,701 & \$35,701 & \$35,701 & \$35,701 & 0.509 & tl. & 42,517 & 0.251 & 77.31~+~0.250 & 1.000 \\
& & (\textbf{+0.68\%}) &  & (\textbf{0.00\%}) & (\textbf{0.00\%}) & (\textbf{0.00\%}) & (\textbf{0.00\%}) & & & & & &  \\

\texttt{240-pserc-api} & \$4,640,589 & ns. & ns. & ns. & \$4,627,635 & \$4,627,155 & ns. & tl. & tl. & tl. & 9.168 & 177.6~+~7.176 & tl. \\
& &  &  &  & (\textbf{--0.28\%}) & (\textbf{--0.29\%}) &  & & & & & &  \\

\texttt{300-ieee-api} & \$684,985 & Inf. & ns. & ns. & \$686,067 & \$684,985 & \$683,968 & 3,399 & tl. & tl. & 5.736 & 135.2~+~5.211 & 35,248 \\
& &  &  &  & (\textbf{+0.16\%}) & (\textbf{0.00\%}) & (\textbf{--0.15\%}) & & & & & &  \\

\texttt{500-goc-api} & \$692,407 & ns. & ns. & ns. & Inf. & \$692,271 & ns. & tl. & tl. & tl. & 21.24 & 340.5 +~23.92 & tl. \\
& &  &  &  &  & (\textbf{--0.02\%}) &  & & & & & &  \\

\texttt{1354-pegase-api} & \$1,498,271 & ns. & ns. & ns. & \$1,496,750 & \$1,496,750 & ns. & tl. & tl. & tl. & 64.28 & 1,030~+~65.63 & tl. \\
& &  &  &  & (\textbf{--0.10\%}) & (\textbf{--0.10\%}) &  & & & & & &  \\
\bottomrule
\end{tabular}
\begin{tablenotes}
\item[*] \scriptsize The percentage differences are calculated as \((\text{Method Cost} - \text{AC-OPF Cost}) / \text{AC-OPF Cost} \times 100\%\).
\item[*] \scriptsize DC-OTS, LPAC-OTS~\cite{Carleton2014}, QC-OTS~\cite{coffrin2015qc}, and AC-OTS are solved using \texttt{PowerModels.jl}.

\item[*] \scriptsize ``Inf.'': Infeasible solution; ``ns.'': Not solved within time limit; ``scb.'': Solution cannot be built (the solver failed to produce a feasible or optimal solution); ``tl.'': Time limit reached (12 hours).
\item[*] \scriptsize The O-DC-OTS time is presented as two values: the time spent on optimizing parameters and the time for solving the DC-OTS problem.

\end{tablenotes}
\end{threeparttable}
\end{adjustbox}
\label{table:cost_timing_comparison}
\end{table*}

To evaluate the efficacy of our proposed ``optimized DC-OTS'' (O-DC-OTS), we conducted extensive numerical comparisons against traditional DC-OTS (i.e., Model~\ref{model:DCOPFOTS_matrixformat} without the $c^{prof}\mathbf{1}^{\top}\mathbf{u}$ term in the objective function), LPAC-OTS~\cite{Carleton2014}, QC-OTS~\cite{coffrin2015qc}, and C-DC-OTS (i.e., Model~\ref{model:DCOPFOTS_matrixformat}) models. We selected a diverse set of test systems from the PGLib-OPF archive~\cite{pglib}, specifically the cases listed in Tables~\ref{table:opened_lines_methods} and~\ref{table:cost_timing_comparison}. These test systems vary significantly in size and complexity, ranging from a small \texttt{3-bus} system to a large-scale \texttt{1354-bus} network, thereby assessing our algorithm's performance across different network scales.
\textcolor{black}{Furthermore, our results indicate that while the inclusion of the switching penalty \(c^{prof}\mathbf{1}^{\top}\mathbf{u}\) (as in C-DC-OTS) improves AC feasibility relative to traditional DC-OTS, the combination with optimized parameters \((\boldsymbol{b}, \boldsymbol{\psi})\) in O-DC-OTS further enhances both operational cost reduction and AC feasibility. We also evaluated the effect of varying the optimization algorithm by testing the Conjugate Gradient method, the Broyden–Fletcher–Goldfarb–Shanno (BFGS) algorithm, the Limited-memory BFGS (L-BFGS) algorithm, and the Truncated Newton Conjugate-Gradient (TNC) method, with results demonstrating the Conjugate Gradient method's superiority.}

All computational experiments were performed on a high-performance computing node equipped with 24 cores and 16 GB of RAM, provided by Georgia Tech's Partnership for an Advanced Computing Environment (PACE). Our algorithm was implemented in PyTorch, and we utilized \texttt{cvxpylayers} \cite{cvxpylayers2019} to compute the necessary derivatives of the loss function with respect to the parameters—specifically, $\frac{\partial \mathcal{L}}{\partial \mathbf{b}}$ and $\frac{\partial \mathcal{L}}{\partial \boldsymbol{\psi}}$. We used the CG implementation from the \texttt{scipy.optimize.minimize} library to minimize the loss function~\eqref{eq:loss_function} based on the sensitivities ${\mathbf{g}^{b}}$ and ${\mathbf{g}^{\psi}}$. For solving the AC-OPF problems and other OTS formulations, we used \texttt{PowerModels.jl}\cite{coffrin2018powermodels}. We used a 12-hour time limit. \textcolor{black}{MILP problems were solved using the solver Gurobi v11.0.3 with default parameter settings and optimality gap of $0.01\%$.}
\textcolor{black}{The switching‑penalty parameter \(c^{\mathrm{prof}}\) is set via a one‑time tuning per test system: beginning at 1\% of the base‑case DC‑OTS cost, we evaluate a few candidate absolute values and choose the penalty that works best for that test case. The resulting \(c^{\mathrm{prof}}\) varies across systems but remains close to the initial estimate.}

\subsection{Results and Analysis}

In this section, we present a comprehensive comparison of our optimized DC-OPF model (O-DC-OTS) against traditional DC-OTS and other advanced OTS formulations such as LPAC-OTS, QC-OTS, and AC-OTS. All evaluations were conducted within the OTS framework using various test cases shown in Tables~\ref{table:opened_lines_methods} and~\ref{table:cost_timing_comparison}. The AC-OPF cost without any line switching serves as the baseline for cost comparisons.

To solve the AC-OTS problem, we used the Juniper solver~\cite{kroger2018juniper}, which is designed for non-convex MINLP problems. Juniper employs a branch-and-bound algorithm combined with local nonlinear programming solvers to handle the continuous relaxation at each node. Due to the non-convexity of the AC power flow equations and the presence of binary variables for line switching, Juniper does not guarantee global optimality. However, it often finds high-quality feasible OTS solutions under AC constraints. Thus, for the small- to moderate-size systems that Juniper can solve, it provides a good proxy for ``best achievable'' solution quality.\footnote{\textcolor{black}{To assess the quality of Juniper’s AC‑OTS solutions, we performed a \emph{bounded exhaustive search} on several small test cases (\texttt{3‑lmbd‑api}, \texttt{5‑pjm‑api}, \texttt{14‑ieee‑api}, \texttt{24‑ieee‑rts‑api}, \texttt{30‑as‑api}, and \texttt{30‑ieee‑api}).  In each case, Juniper’s solution matched the best objective found by enumeration.  For the \texttt{39‑epri‑api} system, enumeration—limited to at most six line openings (the largest number Juniper ever used)—revealed a slightly better configuration (opening lines $4$ and $6$) with cost $\$246,561$ versus Juniper’s $\$246,723$, a $0.07\%$ improvement.}
}

Table~\ref{table:opened_lines_methods} lists the lines opened by each method across different test cases, and Table~\ref{table:cost_timing_comparison} provides operational costs and computation times for each model. The AC-OPF cost without switching is provided as a reference, and the switching decisions from different OTS models were validated by running AC-OPF problems with topologies fixed to the outputs of each model to assess their feasibility and effectiveness.

Our analysis reveals that the traditional DC-OTS model often produces infeasible solutions when validated using the AC-OPF model, primarily due to inaccuracies in capturing voltage violations and line overloads. For example, in the \texttt{14-ieee-api} case, the DC-OTS model recommends opening lines $12$, $13$, and $14$, leading to infeasibility. Likewise, the LPAC-OTS model recommends opening lines $11$ and $20$,  which also leads to infeasibility. Similarly, in the \texttt{73-ieee-rts-api} case, the DC-OTS solution's opening of a large number of lines disrupts network connectivity, resulting in an infeasibility under AC validation.

In contrast, our O-DC-OTS model effectively addresses these shortcomings by optimizing line susceptances ($\mathbf{b}$) and bias parameters ($\boldsymbol{\psi}$), providing a more accurate approximation of AC power flows. In the \texttt{73-ieee-rts-api} case, our model suggests opening only lines $9$ and $50$, resulting in an operational cost of $\$413,133$, which is feasible under AC validation and closer to the AC-OTS cost of $\$385,194$. Compared to the AC-OPF baseline cost of $\$422,627$, this represents a cost reduction of $2.3\%$. While the LPAC-OTS model could not be solved within the time limit, the QC-OTS model achieves a small cost improvement of $1.44\%$.

\subsection{Case Studies}

To provide deeper insights into the performance of our algorithm, we present detailed analyses of selected test cases.

\subsubsection{\texttt{30-as-api} Case}

In the \texttt{30-as-api} test case, the original AC-OPF cost was $\$4,996$. Our proposed O-DC-OTS model reduced this cost to $\$2,797$, representing a cost reduction of $44$\% compared to the baseline. Notably, the LPAC-OTS and QC-OTS models resulted in costs of $\$2,809$ and $\$4,996$, respectively, corresponding to cost reductions of $43.78\%$ and $0\%$. Thus, our algorithm achieves lower costs while maintaining AC feasibility.

\subsubsection{\texttt{39-epri-api} Case}

In the \texttt{39-epri-api} test case, the traditional DC-OTS model produced infeasible solutions under AC validation. The DC-OTS model suggested opening lines $17$, $22$, $31$, and $43$, which makes the AC-OPF problem infeasible. Our O-DC-OTS model, however, recommended opening lines $4$ and $13$, resulting in an operational cost of $\$246,850$, representing a $1.13$\% cost reduction over original AC-OPF cost of $\$249,672$. 
Meanwhile, the decision from the LPAC-OTS model rendered the AC-OPF infeasible, and the QC-OTS model did not lead to any cost improvement.

\subsubsection{\texttt{60-c-api} Case}

In the \texttt{60-c-api} case, both the traditional DC-OTS and our optimized O-DC-OTS models produced feasible solutions when validated under AC-OPF. However, our O-DC-OTS model achieved a lower operational cost of $\$182,028$, compared to $\$183,302$ for the traditional DC-OTS and $\$185,239$ for the original AC-OPF solution. This represents a $0.7\%$ cost reduction relative to the traditional DC-OTS model and a $1.7\%$ improvement over the original AC-OPF solution. These results show that our algorithm can yield more cost-effective transmission switching decisions even when the traditional DC-OTS is feasible. Also note that while the switching decisions from the QC-OTS model were feasible, they led to an increase in the cost function by $0.63\%$.

\subsubsection{\texttt{89-pegase-api} Case}

In the \texttt{89-pegase-api} test case, the LPAC-OTS model could not be constructed, and QC-OTS did not solve within the time limit. On the other hand, \mbox{C-DC-OTS} and O-DC-OTS models identified a set of lines to open (lines $9$ and $34$), resulting in an operational cost of $\$100,702$. This is a $22.64$\% cost reduction compared to the original AC-OPF cost of $\$130,175$ and closely approaches the AC-OTS cost of $\$100,344$ ($22.92\%$ cost reduction).

\subsection{Computational Efficiency}

The computation times reported in Table~\ref{table:cost_timing_comparison} show that our \mbox{O-DC-OTS} model offers a favorable trade-off between accuracy and computational efficiency. While applying a nonlinear branch-and-bound solver like Juniper~\cite{kroger2018juniper} to directly handle the AC-OTS problem as an MINLP provides the most accurate results for small- to moderate-size systems, this scales poorly with increasing system size. By using an MILP formulation, our optimized model significantly reduces computation time, making it more suitable for larger systems.

For very small test cases, the computation time of our \mbox{O-DC-OTS} model is comparable to that of AC-OTS with Juniper~\cite{kroger2018juniper}, typically taking only a few seconds. However, as the system size increases, the computational speed advantages of our model become more apparent. For test cases with several tens of buses or more, our approach demonstrates significant efficiency gains. As an example, for the \texttt{89-pegase-api} system, our \mbox{O-DC-OTS} model provides two orders of magnitude speed improvement relative to AC-OTS with Juniper ($23.47$ seconds, with $22.88$ seconds for training and $0.598$ seconds to solve compared to $1,866$ seconds). In addition to the significant time savings, our model achieved a $22.6\%$ cost reduction compared to the original AC-OPF cost, closely matching the AC-OTS solution from Juniper. This combination of high accuracy and substantially reduced computational burden makes our \mbox{O-DC-OTS} model particularly advantageous for large-scale power systems.

\subsection{Discussion}

Our results demonstrate that the O-DC-OTS significantly enhances both solution quality and computational efficiency. 
%
On average across the test cases where Juniper solved the AC-OTS problem ($14$ out of $18$ test cases), the O-DC-OTS model achieved a cost reduction of approximately $7.7\%$ compared to the original AC-OPF solution. This closely approaches the $8.8\%$ reduction achieved by the AC-OTS model solved with Juniper.
In some cases like \texttt{30-as-api} and \texttt{89-pegase-api}, the \mbox{O-DC-OTS} model achieved even more substantial cost reductions of $44\%$ and $22.6\%$, respectively.

In terms of computational efficiency, the O-DC-OTS model was, on average, approximately $29$ times faster than the AC-OTS model with Juniper across the test cases where both methods solved. This speed advantage is more pronounced in larger systems; e.g., solution times for the \texttt{300-ieee-api} test case were $140$ seconds for \mbox{O-DC-OTS} compared to over $35,000$ seconds for AC-OTS with Juniper (a factor of $251$ speed improvement). Comparing to other methods, the QC-OTS model was, on average, approximately $149$ times slower than O-DC-OTS across the test cases where QC-OTS provided solutions. The LPAC-OTS model often failed to solve larger test cases within reasonable time frames.
Our O-DC-OTS model also demonstrated superior reliability in finding AC-feasible solutions. Among the test cases with more than $60$ buses, LPAC-OTS, QC-OTS, and AC-OTS with Juniper provided solutions for $0\%$, $22\%$, and $55\%$ of the test cases, respectively, while O-DC-OTS successfully solved all of them.


\textcolor{black}{Compared to the C-DC-OTS formulation, it is important to note that while the parameter optimization phase adds to the overall solution time, this additional computational effort can yield substantial improvements in solution accuracy and AC feasibility. The increase in computational time is more than compensated  by the model's ability to reliably produce high-quality, cost-effective, and operationally viable solutions, thereby reinforcing its suitability for real-time OTS applications. We note that our proof‑of‑concept implementation employs a Python‑based gradient optimizer, whereas the DC‑OTS subproblem leverages Gurobi’s highly tuned MILP engine. By parallelizing gradient evaluations and implementing a specialized, parallel CG solver in a low-level language—while also further exploring alternatives such as L‑BFGS and TNC—we anticipate that tuning runtimes could be significantly reduced. Consequently, total solution times are projected to remain within relevant real‑time OTS windows, ensuring both high solution quality and operational viability.}

In summary, O-DC-OTS effectively balances solution quality and computational speed, achieving significant operational cost savings comparable to AC-OTS with Juniper but with much lower computational requirements. Moreover, O-DC-OTS frequently outperforms LPAC-OTS and QC-OTS in terms of both solution quality and computational speed.

%% file: main.bbl
\begin{thebibliography}{10}
\providecommand{\url}[1]{#1}
\csname url@samestyle\endcsname
\providecommand{\newblock}{\relax}
\providecommand{\bibinfo}[2]{#2}
\providecommand{\BIBentrySTDinterwordspacing}{\spaceskip=0pt\relax}
\providecommand{\BIBentryALTinterwordstretchfactor}{4}
\providecommand{\BIBentryALTinterwordspacing}{\spaceskip=\fontdimen2\font plus
\BIBentryALTinterwordstretchfactor\fontdimen3\font minus \fontdimen4\font\relax}
\providecommand{\BIBforeignlanguage}[2]{{%
\expandafter\ifx\csname l@#1\endcsname\relax
\typeout{** WARNING: IEEEtran.bst: No hyphenation pattern has been}%
\typeout{** loaded for the language `#1'. Using the pattern for}%
\typeout{** the default language instead.}%
\else
\language=\csname l@#1\endcsname
\fi
#2}}
\providecommand{\BIBdecl}{\relax}
\BIBdecl

\bibitem{koglin1981first}
H.~Koglin and H.~Muller, ``First experiences with computer-aided corrective switching,'' in \emph{7th Power Systems Computation Conference (PSCC)}, 1981, pp. 474--481.

\bibitem{koglin1982corrective}
H.~Koglin and H.~M{\"u}ller, ``Corrective switching: A new dimension in optimal load flow,'' \emph{International Journal of Electrical Power \& Energy Systems}, vol.~4, no.~2, pp. 142--149, 1982.

\bibitem{cohen1991paradoxical}
J.~E. Cohen and P.~Horowitz, ``Paradoxical behaviour of mechanical and electrical networks,'' \emph{Nature}, vol. 352, no. 6337, pp. 699--701, 1991.

\bibitem{witthaut2012braess}
D.~Witthaut and M.~Timme, ``Braess's paradox in oscillator networks, desynchronization and power outage,'' \emph{New Journal of Physics}, vol.~14, no.~8, p. 083036, 2012.

\bibitem{o2005dispatchable}
R.~P. O'Neill, R.~Baldick, U.~Helman, M.~H. Rothkopf, and W.~Stewart, ``Dispatchable transmission in {RTO} markets,'' \emph{IEEE Transactions on Power Systems}, vol.~20, no.~1, pp. 171--179, 2005.

\bibitem{fisher2008optimal}
E.~B. Fisher, R.~P. O'Neill, and M.~C. Ferris, ``Optimal transmission switching,'' \emph{IEEE Transactions on Power Systems}, vol.~23, no.~3, pp. 1346--1355, 2008.

\bibitem{Shahidehpour2002market}
M.~Shahidehpour, H.~Yamin, and Z.~Y. Li, \emph{Market Operations in Electric Power Systems}.\hskip 1em plus 0.5em minus 0.4em\relax New York: Wiley, 2002.

\bibitem{Conejo2006decomposition}
A.~J. Conejo, E.~Castillo, R.~M{\'\i}nguez, and R.~Garc{\'\i}a-Bertrand, \emph{Decomposition Techniques in Mathematical Programming}.\hskip 1em plus 0.5em minus 0.4em\relax New York: Springer, 2006.

\bibitem{Bacher1986network}
R.~Bacher and H.~Glavitsch, ``Network topology optimization with security constraints,'' \emph{IEEE Transactions on Power Systems}, vol.~1, no.~4, pp. 103--111, Nov. 1986.

\bibitem{granelli2006optimal}
G.~Granelli, M.~Montagna, F.~Zanellini, P.~Bresesti, R.~Vailati, and M.~Innorta, ``Optimal network reconfiguration for congestion management by deterministic and genetic algorithms,'' \emph{Electric Power Systems Research}, vol.~76, no. 6-7, pp. 549--556, 2006.

\bibitem{Schnyder1988integrated}
G.~Schnyder and H.~Glavitsch, ``Integrated security control using an optimal power flow and switching concepts,'' \emph{IEEE Transactions on Power Systems}, vol.~3, no.~2, pp. 782--790, May 1988.

\bibitem{Rolim1999corrective}
J.~G. Rolim and L.~J.~B. Machado, ``A study of the use of corrective switching in transmission systems,'' \emph{IEEE Transactions on Power Systems}, vol.~14, no.~1, pp. 336--341, Feb. 1999.

\bibitem{hedman2008optimal}
K.~W. Hedman, R.~P. O'Neill, E.~B. Fisher, and S.~S. Oren, ``Optimal transmission switching--{S}ensitivity analysis and extensions,'' \emph{IEEE Transactions on Power Systems}, vol.~23, no.~3, pp. 1469--1479, 2008.

\bibitem{Khodaei2010}
A.~Khodaei, M.~Shahidehpour, and S.~Kamalinia, ``Transmission switching in expansion planning,'' \emph{IEEE Transactions on Power Systems}, vol.~25, no.~3, pp. 1722--1733, 2010.

\bibitem{capitanescu2011state}
F.~Capitanescu, T.~Van~Cutsem, L.~Wehenkel, M.~Glavic, F.~Marinho, F.~Torres, and J.~Martinez, ``State-of-the-art, challenges, and future trends in security constrained optimal power flow,'' \emph{Electric Power Systems Research}, vol.~81, no.~8, pp. 1731--1741, 2011.

\bibitem{stott2009dc}
B.~Stott, J.~Jardim, and O.~Alsa{\c{c}}, ``{DC} power flow revisited,'' \emph{IEEE Transactions on Power Systems}, vol.~24, no.~3, pp. 1290--1300, 2009.

\bibitem{lehmann2014complexity}
K.~Lehmann, A.~Grastien, and P.~Van~Hentenryck, ``The complexity of {DC}-switching problems,'' \emph{arXiv:1411.4369}, 2014.

\bibitem{ruiz2011fast}
P.~A. Ruiz, J.~M. Foster, A.~Rudkevich, and M.~C. Caramanis, ``On fast transmission topology control heuristics,'' in \emph{IEEE Power and Energy Society General Meeting}, 2011.

\bibitem{ruiz2012tractable}
------, ``Tractable transmission topology control using sensitivity analysis,'' \emph{IEEE Transactions on Power Systems}, vol.~27, no.~3, pp. 1550--1559, 2012.

\bibitem{kocuk2016cycle}
B.~Kocuk, H.~Jeon, S.~S. Dey, J.~Linderoth, J.~Luedtke, and X.~A. Sun, ``A cycle-based formulation and valid inequalities for {DC} power transmission problems with switching,'' \emph{Operations Research}, vol.~64, no.~4, pp. 922--938, 2016.

\bibitem{hijazi2017convex}
H.~Hijazi, C.~Coffrin, and P.~V. Hentenryck, ``Convex quadratic relaxations for mixed-integer nonlinear programs in power systems,'' \emph{Mathematical Programming Computation}, vol.~9, pp. 321--367, 2017.

\bibitem{fattahi2018bound}
S.~Fattahi, J.~Lavaei, and A.~Atamt{\"u}rk, ``A bound strengthening method for optimal transmission switching in power systems,'' \emph{IEEE Transactions on Power Systems}, vol.~34, no.~1, pp. 280--291, 2018.

\bibitem{dey2022node}
S.~S. Dey, B.~Kocuk, and N.~Redder, ``Node-based valid inequalities for the optimal transmission switching problem,'' \emph{Discrete Optimization}, vol.~43, p. 100683, 2022.

\bibitem{barrows2014}
C.~Barrows, S.~Blumsack, and P.~Hines, ``Correcting optimal transmission switching for {AC} power flows,'' in \emph{47th Hawaii International Conference on System Sciences (HICSS)}, January 2014, pp. 2374--2379.

\bibitem{coffrin2014switching}
C.~Coffrin, H.~L. Hijazi, K.~Lehmann, and P.~Van~Hentenryck, ``Primal and dual bounds for optimal transmission switching,'' in \emph{18th Power Systems Computation Conference (PSCC)}, June 2014.

\bibitem{jabr2006radial}
R.~A. Jabr, ``Radial distribution load flow using conic programming,'' \emph{IEEE Transactions on Power Systems}, vol.~21, no.~3, pp. 1458--1459, 2006.

\bibitem{coffrin2015qc}
C.~Coffrin, H.~L. Hijazi, and P.~Van~Hentenryck, ``The {QC} relaxation: A theoretical and computational study on optimal power flow,'' \emph{IEEE Transactions on Power Systems}, vol.~31, no.~4, pp. 3008--3018, 2015.

\bibitem{bai2008semidefinite}
X.~Bai, H.~Wei, K.~Fujisawa, and Y.~Wang, ``Semidefinite programming for optimal power flow problems,'' \emph{International Journal of Electrical Power \& Energy Systems}, vol.~30, no. 6-7, pp. 383--392, 2008.

\bibitem{Lavaei2012}
J.~Lavaei and S.~H. Low, ``Zero duality gap in optimal power flow problem,'' \emph{IEEE Transactions on Power Systems}, vol.~27, no.~1, pp. 92--107, 2012.

\bibitem{molzahn2013implementation}
D.~K. Molzahn, J.~T. Holzer, B.~C. Lesieutre, and C.~L. DeMarco, ``Implementation of a large-scale optimal power flow solver based on semidefinite programming,'' \emph{IEEE Transactions on Power Systems}, vol.~28, no.~4, pp. 3987--3998, 2013.

\bibitem{Carleton2014}
C.~Coffrin and P.~Van~Hentenryck, ``A linear-programming approximation of {AC} power flows,'' \emph{INFORMS Journal on Computing}, vol.~26, no.~4, pp. 718--734, 2014.

\bibitem{guo2022tightening}
C.~Guo, H.~Nagarajan, and M.~Bodur, ``Tightening quadratic convex relaxations for the {AC} optimal transmission switching problem,'' \emph{arXiv:2212.12097}, 2022.

\bibitem{kocuk2017new}
B.~Kocuk, S.~S. Dey, and X.~A. Sun, ``New formulation and strong {MISOCP} relaxations for {AC} optimal transmission switching problem,'' \emph{IEEE Transactions on Power Systems}, vol.~32, no.~6, pp. 4161--4170, 2017.

\bibitem{bestuzheva2020}
K.~Bestuzheva, H.~Hijazi, and C.~Coffrin, ``Convex relaxations for quadratic on/off constraints and applications to optimal transmission switching,'' \emph{INFORMS Journal on Computing}, vol.~32, no.~3, pp. 682--696, 2020.

\bibitem{taheri2023optimizing}
B.~Taheri and D.~K. Molzahn, ``{Optimizing parameters of the DC power flow},'' \emph{Electric Power Systems Research}, vol. 235, no. 110719, October 2024, {\rm presented at the} \textit{23rd Power Systems Computation Conference (PSCC)}.

\bibitem{taheri2024DCOPF}
------, ``Improving the accuracy of {DC} optimal power flow formulations via parameter optimization,'' \emph{arXiv:2410.11725}, 2024.

\bibitem{fiacco1990nonlinear}
A.~V. Fiacco and G.~P. McCormick, \emph{Nonlinear Programming: Sequential Unconstrained Minimization Techniques}.\hskip 1em plus 0.5em minus 0.4em\relax SIAM, 1990.

\bibitem{robinson1980strongly}
S.~M. Robinson, ``Strongly regular generalized equations,'' \emph{Mathematics of Operations Research}, vol.~5, no.~1, pp. 43--62, 1980.

\bibitem{amos2017optnet}
B.~Amos and J.~Z. Kolter, ``Optnet: Differentiable optimization as a layer in neural networks,'' in \emph{International Conference on Machine Learning (ICML)}.\hskip 1em plus 0.5em minus 0.4em\relax PMLR, 2017, pp. 136--145.

\bibitem{cvxpylayers2019}
A.~Agrawal, B.~Amos, S.~Barratt, S.~Boyd, S.~Diamond, and Z.~Kolter, ``Differentiable convex optimization layers,'' in \emph{Advances in Neural Information Processing Systems (NeurIPS)}, 2019.

\bibitem{karush1939minima}
W.~Karush, ``Minima of functions of several variables with inequalities as side constraints,'' \emph{M. Sc. Dissertation. Dept. of Mathematics, Univ. of Chicago}, 1939.

\bibitem{kuhn1951proceedings}
H.~Kuhn and A.~Tucker, ``Nonlinear programming,'' in \emph{2nd Berkeley Symposium}, 1951, pp. 481--492.

\bibitem{nocedal2006numerical}
J.~Nocedal and S.~Wright, \emph{Numerical Optimization}.\hskip 1em plus 0.5em minus 0.4em\relax New York, NY: Springer Science \& Business Media, 2006.

\bibitem{pglib}
{IEEE PES Task Force on Benchmarks for Validation of Emerging Power System Algorithms}, ``The {P}ower {G}rid {L}ibrary for benchmarking {AC} optimal power flow algorithms,'' August 2019, {\it arXiv:1908.02788}.

\bibitem{coffrin2018powermodels}
C.~Coffrin, R.~Bent, K.~Sundar, Y.~Ng, and M.~Lubin, ``{PowerModels.jl}: An open-source framework for exploring power flow formulations,'' in \emph{20th Power Systems Computation Conference (PSCC)}, 2018.

\bibitem{kroger2018juniper}
O.~Kr{\"o}ger, C.~Coffrin, H.~Hijazi, and H.~Nagarajan, ``Juniper: An open-source nonlinear branch-and-bound solver in {Julia},'' in \emph{15th International Conference on the Integration of Constraint Programming, Artificial Intelligence, and Operations Research (CPAIOR)}, 2018, pp. 377--386.

\bibitem{han2023optimal}
T.~Han, D.~J. Hill, and Y.~Song, ``Optimal transmission switching with uncertainties from both renewable energy and {N-k} contingencies,'' \emph{IEEE Transactions on Sustainable Energy}, vol.~14, no.~4, pp. 1964--1978, 2023.

\bibitem{roald2022review}
L.~A. Roald, D.~Pozo, A.~Papavasiliou, D.~K. Molzahn, J.~Kazempour, and A.~Conejo, ``Power systems optimization under uncertainty: A review of methods and applications,'' \emph{Electric Power Systems Resesarch}, vol. 214, p. 108725, 2023, {\rm presented at the} \emph{22nd Power Systems Computation Conference (PSCC)}.

\end{thebibliography}
